\documentclass[usenatbib]{mn2e}
\usepackage{epsfig}
\usepackage{amsmath}
\usepackage{amssymb}
\usepackage{journal_names}
\usepackage{color}
\usepackage{multirow}

\begin{document}

\title[NGC 4365 GC Kinematics]{The SLUGGS Survey: Globular cluster system kinematics and substructure in NGC 4365}
\author[Blom~et~al.~]{Christina Blom$^{1\star}$, Duncan A.\ Forbes$^{1}$, Jean P.\ Brodie$^{2}$, Caroline Foster$^{3}$, \\
\\
\normalfont{\LARGE Aaron J.\ Romanowsky$^{2}$, Lee R.\ Spitler$^{1}$ and Jay Strader$^{4}$} \\
\\
$^1$ Centre for Astrophysics \& Supercomputing, Swinburne University, Hawthorn VIC 3122, Australia\\
$^2$ University of California Observatories, 1156 High St., Santa Cruz, CA 95064, USA\\
$^3$ European Southern Observatory, Alonso de Cordova 3107, Vitacura, Santiago, Chile\\
$^4$ Harvard-Smithsonian Center for Astrophysics, 60 Garden St., Cambridge, MA 02138, USA\\
\\
$^\star$ \normalfont{Email: cblom@astro.swin.edu.au}}

\date{\today}
\maketitle

\begin{abstract}
We present a kinematic analysis of the globular cluster (GC) system of the giant elliptical galaxy NGC 4365 and find several distinct kinematic substructures. This analysis is carried out using radial velocities for 269 GCs, obtained with the DEIMOS instrument on the Keck II telescope as part of the SAGES Legacy Unifying Globulars and Galaxies Survey (SLUGGS). 
We find that each of the three (formerly identified) GC colour subpopulations reveal distinct rotation properties. The rotation of the green GC subpopulation is consistent with the bulk of NGC 4365's stellar light, which `rolls' about the photometric major axis. The blue and red GC subpopulations show `normal' rotation about the minor axis. We also find that the red GC subpopulation is rotationally dominated beyond 2.5 arcmin ($\sim17$ kpc) and that the root mean squared velocity of the green subpopulation declines sharply with radius suggesting a possible bias towards radial orbits relative to the other GC subpopulations. Additionally, we find a population of low velocity GCs that form a linear structure running from the SW to the NE across NGC 4365 which aligns with the recently reported stellar stream towards NGC 4342. These low velocity GCs have $g'-i'$ colours consistent with the overall NGC 4365 GC system but have velocities consistent with the systemic velocity of NGC 4342.  We discuss the possible formation scenarios for the three GC subpopulations as well as the possible origin of the low velocity GC population. 
\vspace{0.3in}
\end{abstract}

\section{Introduction}

\begin{figure*}\centering
 \includegraphics[width=0.85\textwidth]{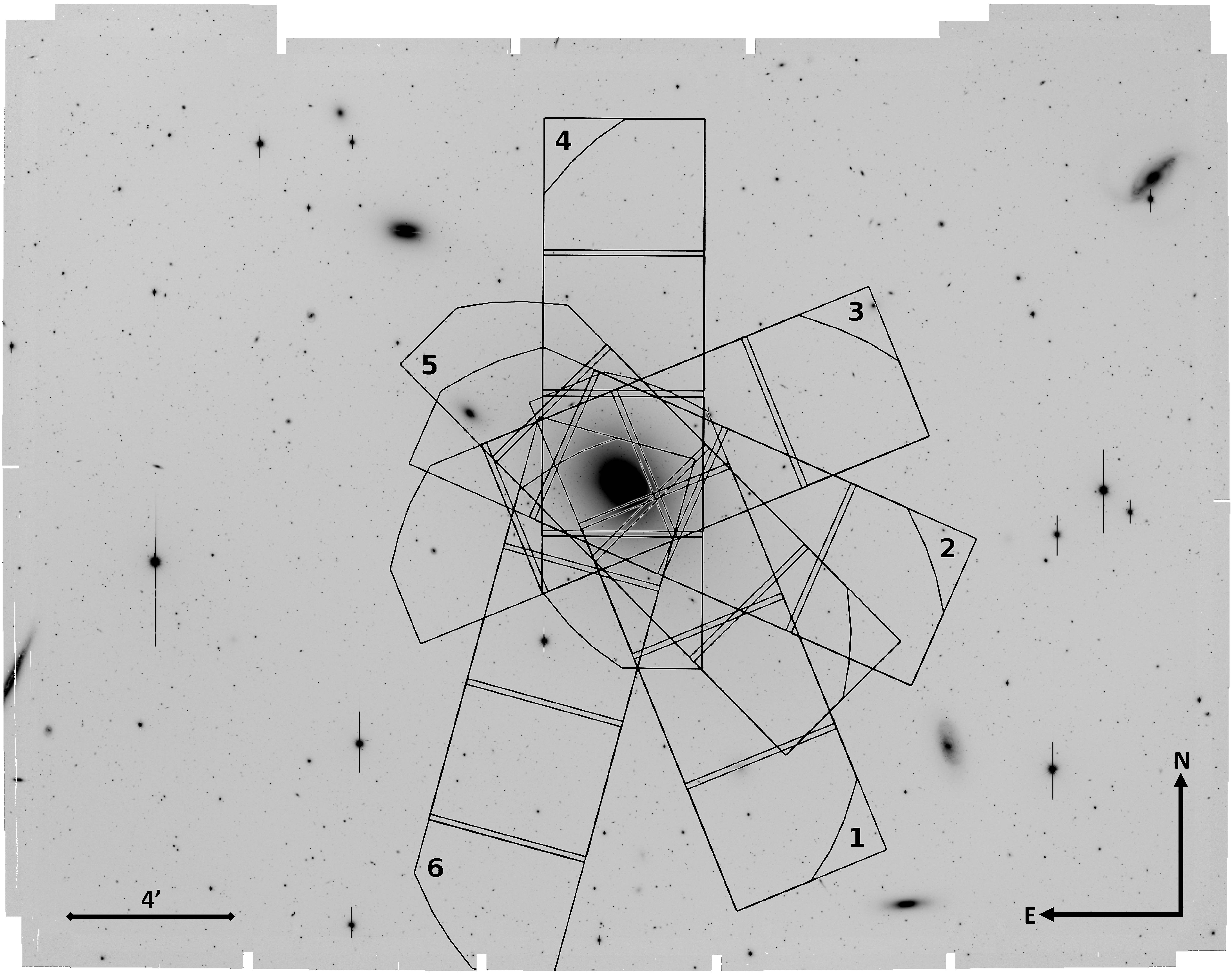}
 \caption{The positions of the six DEIMOS slitmasks plotted on the Subaru Suprime-Cam (S-Cam) $i'$ filter image. The scale is shown in the bottom left corner of the $35 \times 27$ arcmin image. It is centred on $\alpha$=12:24:26.824; $\delta$=+07:19:03.52 (J2000.0). At a distance of $23.1\pm0.8$ Mpc \citep{FCS5} $1\,\mathrm{arcmin} =6.72$ kpc.}
 \label{fig:footprints}
\end{figure*}


Globular clusters (GCs), as some of the oldest and densest stellar systems in the universe, present intriguing questions about their own formation while also being very useful tracers of the formation of their host galaxy \citep{Br06}. They probably formed during violent star formation episodes in their host galaxy (or host galaxy progenitors) and therefore we can use their spatial distribution and kinematic properties to unravel the formation history of galaxies. GCs are more numerous and more easily identifiable in elliptical galaxies (where bright star forming clumps and dust obscuration is minimal) than in spiral galaxies. 

NGC 4365 is a giant elliptical galaxy (E3), $M_B=-21.3$ $M_V=-22.7$ mag \citep{VCS6, Ko09}, with various intriguing and unusual properties as highlighted below. We use these unusual properties of NGC 4365 to constrain galaxy and GC formation scenarios by placing the extra information contained in this system in the overall framework of `normal' giant elliptical galaxies. 

The GC systems of most giant elliptical galaxies are bimodal in optical colours \citep{Br06,VCS9}. There is a growing body of evidence that the relative age difference between GC subpopulations is small \citep{St05} and the distinct colours of the commonly observed subpopulations are due to bimodality in GC metallicity \citep{St07,Wo10,AB11} that are likely explained by two formation mechanisms, sites or epochs. Possible formation scenarios for bimodal metallicity distributions in GC systems are the major merger \citep{Ze93}, multiphase collapse \citep*{F97} and accretion scenarios \citep{C98}. An alternative view is that the bimodal GC colour distributions can be observed from a unimodal GC metallicity distribution, because of strong nonlinearity in the relationship between colour and metallicity \citep{Y06,Ca07,Bl10} and in this case the kinematic properties of each colour subpopulation are not expected to be significantly different.

The GC system of NGC 4365 has three GC subpopulations, that includes an additional subpopulation of `green' GCs between the commonly observed blue and red GC subpopulations. This additional subpopulation, thought at the time to be younger than the `normal' two subpopulations, was discovered by \citet{Pu02} using a combination of near-IR and optical photometry. \citet{Br05} suggested that the three subpopulations might be found as trimodality in optical colours and \citet{La05} found that the green subpopulation was restricted to small galactocentric radii. 

Specifically for NGC 4365, two independent methods have confirmed the GCs for all three subpopulations (blue, green and red) to be older than $\sim10$ Gyr. \citet{Br05} observed a spectroscopic sample of 22 GCs to measure their ages and metallicities using Lick indices from the Low-Resolution Imaging Spectrograph (LRIS) on the Keck telescope. They found that NGC 4365 has three old GC subpopulations with metal poor (blue), intermediate metallicity (green) and metal rich (red) GCs. \citet{Ch11} presented a near-IR photometry and optical analysis of 99 GCs with much higher precision than previously possible, and determined that there is no observable offset in the mean age between the GCs of NGC 4365 and GC populations of other giant ellipticals. This indicates that there is no significant population of young GCs in NGC 4365.

In addition to its almost unique GC system, NGC 4365 has very unusual stellar kinematic properties. The kinematically distinct core (KDC) at its centre is relatively uncommon, seen in $\sim10$ per cent of early type (E and S0) galaxies in the ATLAS$^{\mathrm{3D}}$ volume limited survey \citep{Kr11}.  Also very uncommon is the starlight kinematics outside the KDC which `rolls' about the major axis rather than rotating about the minor axis \citep{Su95}. Less than 5 per cent of galaxies, in the ATLAS$^{3D}$ volume limited survey, show this kinematic behaviour. \citet{Da01} used the SAURON integral field spectrograph to map its kinematic and metallicity structure in two dimensions, finding stars both inside and outside the KDC to be older than 10 Gyr. NGC 4365 is one of only two early type galaxies, in the ATLAS$^{3D}$sample of 260, that have both a KDC and a near $90^\circ$ misalignment between the photometric and kinematic major axes (i.e.\ rolling rather than rotating stars).

Lastly \citet{Bo12}, see also Mihos et al.\ (2012, in prep), presented evidence of a stream of stellar light extending $\sim200$ kpc southwest and $\sim 100$ kpc northeast from NGC 4365. This strong indicator of a very recent $\sim10:1$ merger (occurring with in the last few Gyr) presents a possible cause for the unique properties of NGC 4365.

\citet{Blom12} presented an optical photometry analysis of $\sim 4000$ NGC 4365 GCs, using the most spatially extended sample of GCs analysed to date (observed with Subaru/Suprime-Cam and supplemented with archival imaging from Hubble Space Telescope/Advanced Camera for Surveys) and found different spatial distributions, sizes and mass distributions for the three subpopulations of NGC 4365 GC system. They also concluded that separate formation scenarios are required to explain the existence of the three separate subpopulations of GC in NGC 4365.

Kinematic analysis of the three GC subpopulations in NGC 4365 is key to disentangling the possible formation scenarios of GC systems. Several recent investigations have shown that the kinematic features of the two standard GC subpopulations (seen in other giant elliptical galaxies) are different \citep{Le10a,Ar11,Fos11}. When combined with galaxy formation models this information constrains the possible formation scenarios of GCs. For example, \citet{Fos11} compare kinematics of the GC subpopulations in NGC 4494 with simulations such as those described in \citet{Be05} and conclude that the galaxy has undergone a recent major merger of similar disk galaxies. The addition of the kinematic properties of a third GC subpopulation limits the possible formation scenarios even further.


In Section 2 we first describe the preparation, observation and reduction of the spectroscopic sample of GCs and then investigate the colour/metallicity and line-of-sight velocity distributions of the GCs. The separation of the GCs into subpopulations and analysis of the kinematic features of the three subpopulations are presented in Section 3. We then discuss the results and summarise our conclusions regarding the possible formation scenarios for NGC 4365 in Sections 4 and 5. The Appendices contain the kinematic fits for all the different methods of subpopulation separation we investigated, and an alternative assumption of position angle for kinematic fitting.

\section{Spectroscopic sample}
\subsection{Data acquisition}

\begin{figure}\centering
 \includegraphics[width=0.49\textwidth]{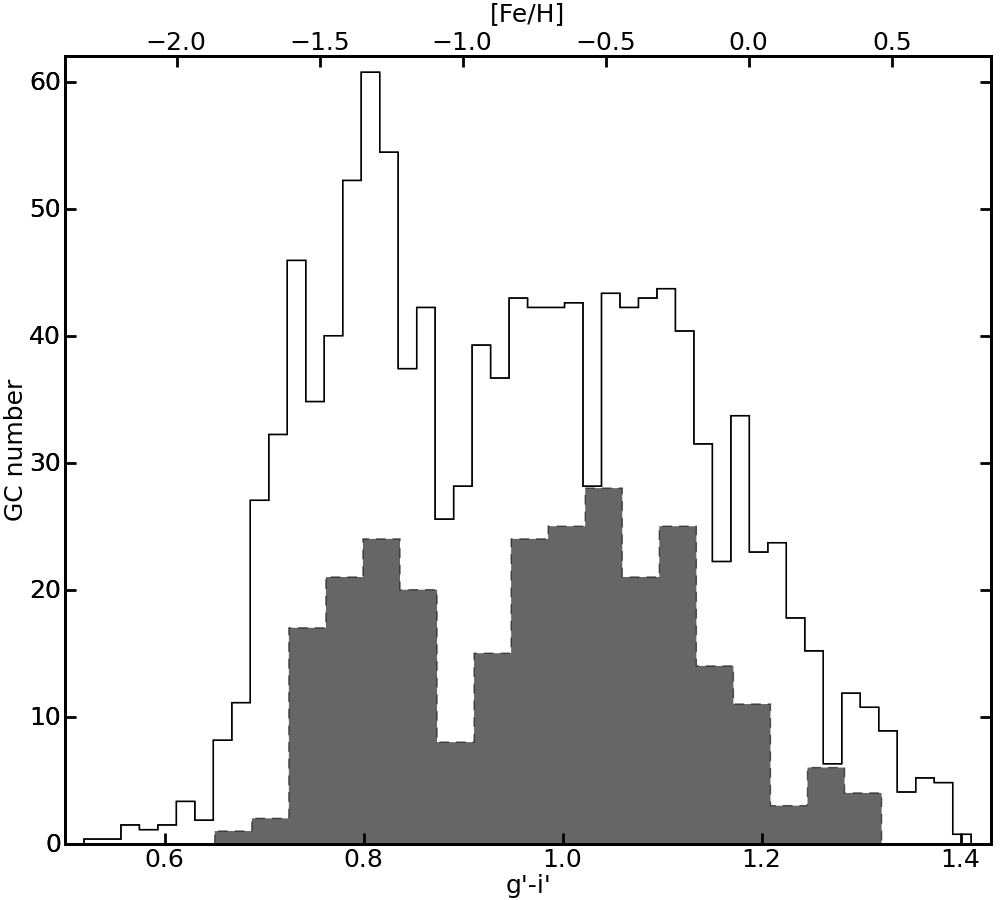}
 \caption{Colour distribution (reddening corrected) of GCs with radial velocity measurements compared with the total GC photometric sample. The grey histogram shows the kinematic sample and the white histogram shows the distribution of photometric candidate GCs brighter than the turnover magnitude, $i'=23.6$, and over all observable radii, $\sim 20$ arcmin, (scaled down by a factor of $2.7$ for comparison purposes). The kinematic sample is a subsample of the photometric candidate sample. The top x-axis shows the [Fe/H] metallicity determined from the empirical conversion published in \citet{Le10b}. We note the three GC subpopulations with peaks at $g'-i'=$ 0.8, 0.97 and 1.13. We also note that the blue GCs in the photometric candidate sample are underrepresented in the kinematic sample with respect to the number of redder GCs ($g'-i'>0.9$).}
 \label{fig:hist}
\end{figure}

The GC spectroscopic sample presented in this work was obtained from six multi-object DEIMOS slitmasks observed in clear dark time on the Keck II telescope. This galaxy was observed as part of the SAGES Legacy Unifying Globulars and Galaxies Survey (SLUGGS\footnote{http://sluggs.swin.edu.au/}) (Brodie et al. 2012 in prep). Four masks were observed on 2010, January 11 and 12 with seeing between 0.6 and 0.9 arcsec and a further two masks were observed on 2010, February 18 with seeing of $\sim0.85''$. \citet{Blom12} used Subaru/Suprime-Cam (S-Cam) and Hubble Space Telescope / Advanced Camera for Surveys (HST/ACS) imaging to identify GC candidates associated with NGC 4365 and publish extinction corrected $g'$, $r'$ and $i'$ magnitudes for S-Cam and/or $g$ and $z$ magnitudes for HST/ACS candidates. A total of 443 GC candidates with $i'$ brighter than $23$ mag were placed in slits, for which we aimed to obtain a spectrum of high enough signal-to-noise (S/N $\sim 5$) to determine a line-of-sight radial velocity. For all six masks a 1200 l mm$^{-1}$ grating was used, with 1 arcsec wide slits and centred on 7800 \AA$\,$ to obtain $\sim$ 1.5 \AA$\,$ resolution. This set up \citep[see also][]{Fos11,Ar11,St11,Ro09,Ro12} allows observations from $\sim$  6550 - 8900 \AA$\,$, which covers the wavelength range in which we expect the red-shifted Calcium II triplet (CaT) absorption features (8498, 8542 and 8662 \AA). It sometimes also includes the red-shifted H$\alpha$ line (6563 \AA) at the blue end of the observed spectrum. 

Footprints of the six DEIMOS masks plotted on the S-Cam $i'$ filter image, used to identify GCs around NGC 4365, are shown in Fig. \ref{fig:footprints}. We obtained total exposure times of 110, 40, 105, 96, 120 and 75 minutes for each of the six masks respectively. For each mask the total exposure time was split into three or four separate exposures to minimize the effects of cosmic rays. The total exposure times for each mask were independently constrained by seeing and object visibility during the night. The median seeing was 0.8, 0.7, 0.75, 0.7, 0.9 and 0.7 arcsec for each of the six masks respectively. Masks two and six were observed at the end of their respective observing nights.

The raw data were reduced to one dimensional spectra using a customised version of the \textsc{DEEP}2 galaxy survey data reduction pipeline \citep[\textsc{IDL SPEC2D;}][]{spec2d1,spec2d2}. The pipeline uses dome flats, NeArKrXe arc lamp spectra and sky light visible in each slit to perform flat fielding, wavelength calibration and local sky subtraction, respectively. It outputs the object spectrum as well as the locally subtracted sky spectrum.

\subsection{Obtaining line-of-sight velocities}
We obtained reliable line-of-sight radial velocities for 252 GCs. This increases the previously available NGC 4365 spectroscopic GC sample \citep[33 individual GCs in][combined]{La03,Br05} by an order of magnitude.

The line-of-sight radial velocity is determined by the peak of the cross correlation function (obtained using the \textsc{IRAF} task \textsc{RV.FXCOR}) between the spectrum and 13 stellar templates. The velocity measurement of a GC is considered reliable if the correlation function has a single, identifiable peak and at least two of the four absorption lines (three CaT and one H$\alpha$) are clearly visible in the red-shifted spectrum. In a few cases one or other of these visual checks was not convincing and the resulting GC velocity is recorded as a borderline velocity measurement (17 objects in total). The final catalogue of 269 radial velocities for NGC 4365 GCs are given in Pota et al.\ (2012, in prep.) 

\begin{figure*}\centering
 \includegraphics[width=0.98\textwidth]{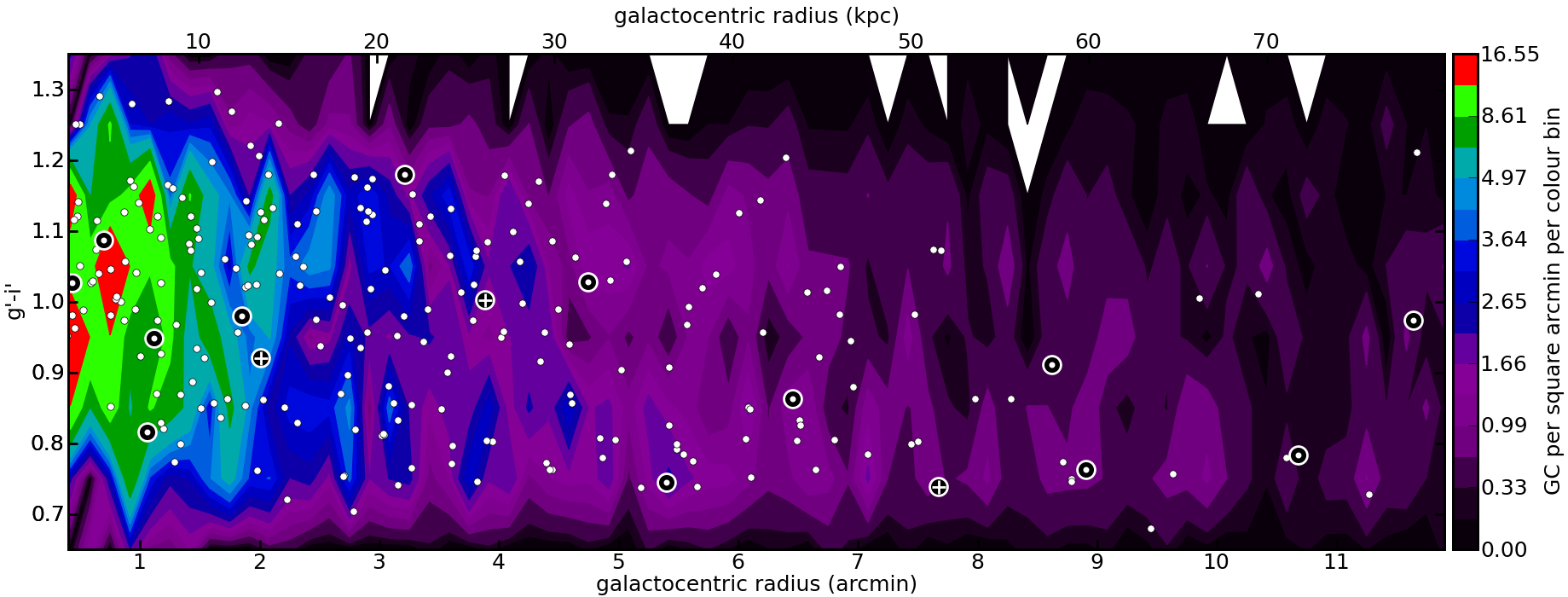}
 \caption{GC surface density with colour and galactocentric radius. The colour scale indicates object number per arcmin$^2$ per colour bin ranging from black at low density to red at high density. The colour-radius position of spectroscopically confirmed GCs are overplotted with white dots. GCs with radial velocities more than $2\sigma$ from the host galaxy are marked with black plusses (positive) and dots (negative). The distribution of GCs with observed line-of-sight radial velocities is consistent with the underlying distribution of photometrically observed GCs in the same radial range and there is no indication that the GCs with $>2\sigma$ low velocities deviate from the underlying distribution either.}
 \label{fig:dens}
\end{figure*}

The colour distribution of the kinematic spectroscopic sample is compared with that of the photometric sample in Fig. \ref{fig:hist}. The photometric sample includes GC candidates with $g'-i'$ colours between 0.5 and 1.4, showing overdensities at $g'-i'\sim0.8$ and between 0.95 and 1.2 as well as an underdensity at $g'-i' \sim 0.9$. Ideally the kinematic sample will have the same colour distribution, scaled down in number, as the photometric sample. In this ideal case we would be able to extrapolate conclusions drawn from the spectroscopic samples to the entire GC population without caveat. Fig. \ref{fig:hist} shows that while we do not completely achieve this ideal, the spectroscopic sample does cover the full colour range and shows a significant underdensity at intermediate colours between the blue and the green subpopulations (seen at $g'-i' \sim 0.9$). In the spectroscopic sample blue GCs are under represented with respect to the green/red GCs compared to the photometric sample. This is because the six DEIMOS pointings have a much smaller radial extent ($\sim 12$ arcmin) than the S-Cam imaging ($\sim 20$ arcmin) and at larger galactocentric radii there are many more blue GCs than red or green GCs \citep[see Fig. \ref{fig:dens} and][]{Blom12}. 

\begin{figure}\centering
 \includegraphics[width=0.49\textwidth]{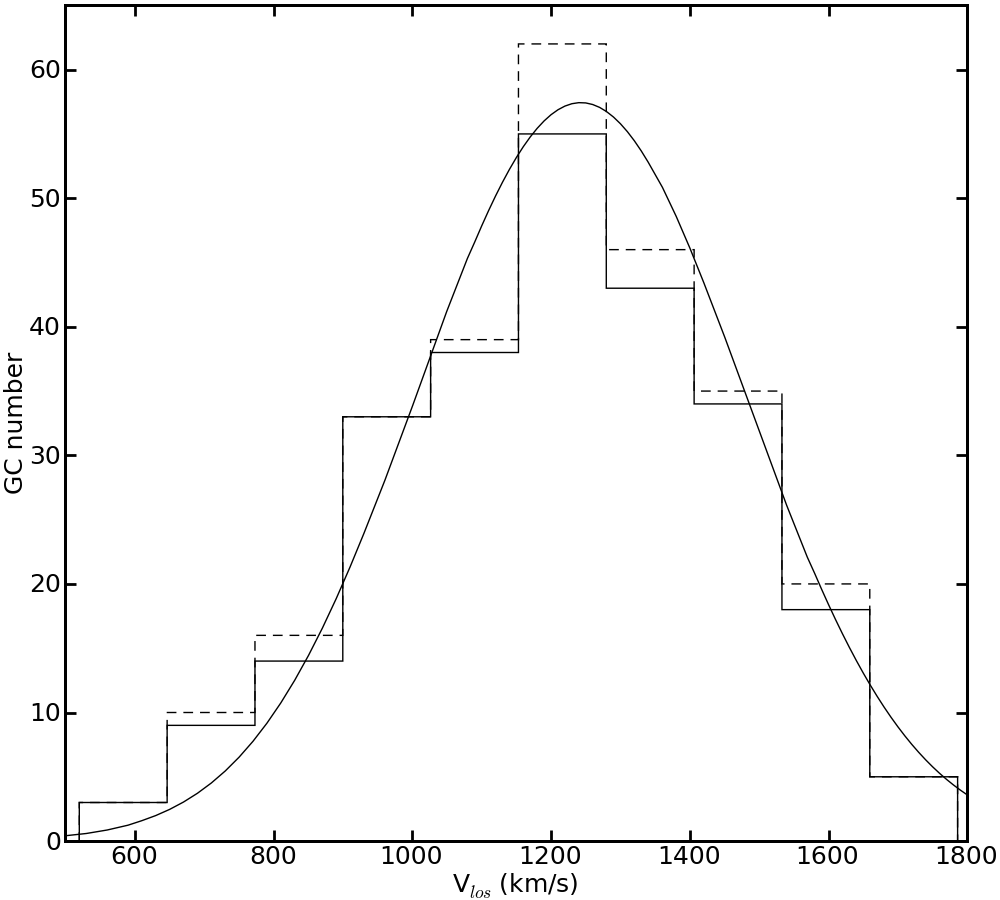}
 \caption{Line-of-sight velocity distribution for GCs in the kinematic sample. The distribution of the secure line-of-sight velocity measurements (252 GCs) is plotted with a solid line and the borderline velocity measurements (17 GCs) are added to the distribution with a dashed line. The plotted Gaussian distribution has a peak at 1243 km s$^{-1}$ and a velocity dispersion of 237 km s$^{-1}$. The velocity distribution of the GCs agrees well with the Gaussian at high velocities but is skewed to low line-of-sight velocities.} 
 \label{fig:vels}
\end{figure}

\begin{figure}\centering
 \includegraphics[width=0.49\textwidth]{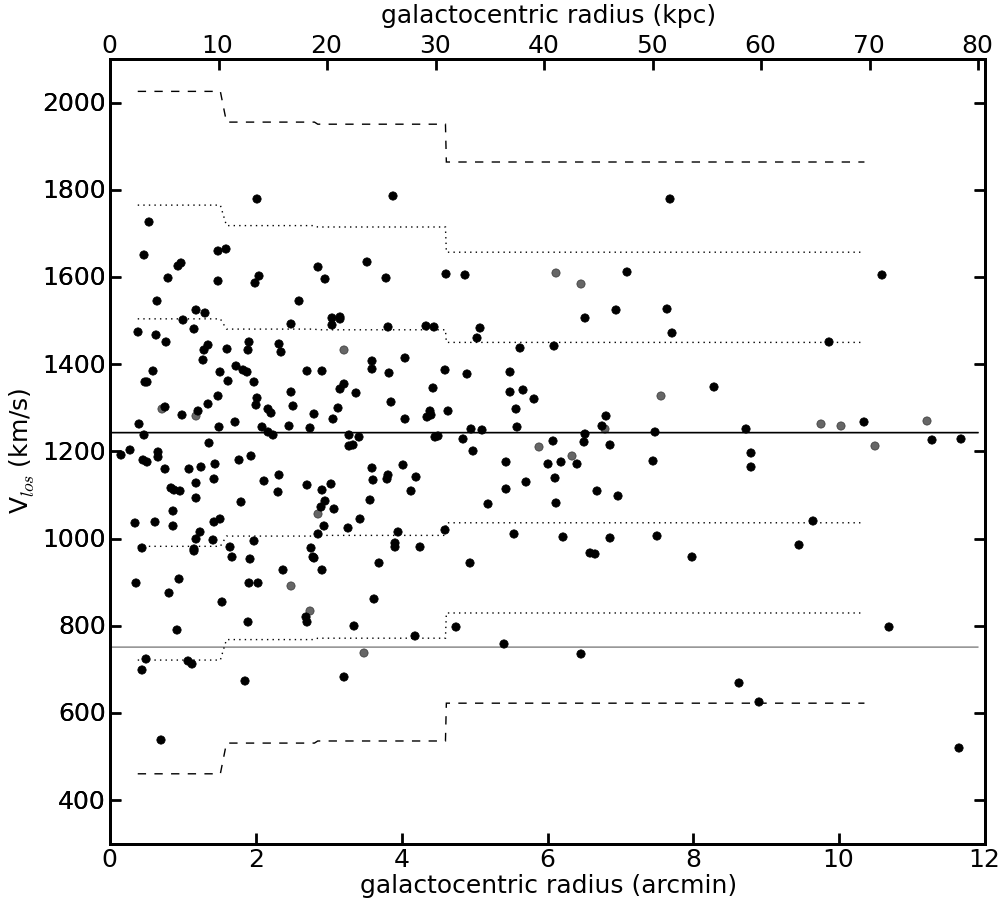}
 \caption{Line-of-sight radial velocity ($V_{los}$) of GCs plotted against galactocentric radius. Filled black circles show the 252 GCs with reliable radial velocity measurements, while grey circles show borderline GC detections. The solid black line is plotted at the galaxy systemic velocity $V_{sys}=1243$ km s$^{-1}$, dotted lines show $1\sigma$ and $2\sigma$ envelopes and the dashed line shows the $3\sigma$ envelope. The $\sigma$ values were calculated in bins of 29 GCs with $V_{los}>1243$ km s$^{-1}$, excluding borderline GCs. Note the large number of low velocity ($V_{los}\sim 700$ km s$^{-1}$) $>2\sigma$ outliers. The solid grey line is plotted at the systemic velocity of the nearby galaxy NGC 4342 \citep[$V_{sys}=751$ km s$^{-1}$,][]{redshifts}. which lies $\sim 35$ arcmin from NGC 4365}
 \label{fig:sigma}
\end{figure}

We expect to see a roughly Gaussian distribution in the line-of-sight velocities of GCs around NGC 4365. In Fig. \ref{fig:vels} we show a Gaussian distribution overplotted on the velocity distribution of NGC 4365's GCs. The peak is set to the literature value for NGC 4365's recession velocity \citep[$v_{sys}=1243$ km s$^{-1}$,][]{vSys} and the standard deviation is set to 237 km s$^{-1}$. This standard deviation was calculated from one side of the velocity distribution, i.e.\ from 68 percent of the GCs with $V_{los}>$ 1243 km s$^{-1}$. The peak agrees well with the observed GC velocity distribution but an excess of GCs is seen above the overplotted Gaussian at low velocities in Fig. \ref{fig:vels}.

We can calculate the skewness ($\xi_3$) of the velocity sample using the standardised third moment \citep{Skew},
\begin{equation}
\xi_3 = \frac{\sqrt{N(N-1)}}{N(N-2)}\frac{\sum_{i=1}^{N}(x_i-\overline{x})^3}{s^3}
\end{equation}
where $N$ is the total number of GCs in the sample, $x_i$ is the line-of-sight velocity of the $i^{th}$ GC, $\overline{x}$ is the mean line-of-sight velocity of the sample and $s$ is the standard deviation of the sample. The calculated skewness is not strongly dependent on whether the entire GCs sample (including borderline velocity measurements) or only the secure velocity measurements are used,
\begin{align}
\xi_{3,all} = -0.20 \notag \\
\xi_{3,secure} = -0.19 \notag
\end{align}
 The standard error in skewness ($\epsilon_{\xi_{3}}$) can be calculated from the number of GCs in the sample \citep{Skew},
\begin{equation}
\epsilon_{\xi_{3}} = \sqrt{\frac{6N(N-1)}{(N-2)(N+1)(N+3)}} = 0.15
\end{equation}
The measured negative skewness of the GC line-of-sight velocity distribution ($-0.19\pm0.15$) is only marginally significant, given the relatively small number of GCs contributing to the skewness. 
We further investigate the possible causes of this negative skewness in the velocity distribution.

\subsection{Low velocity GCs}

We explore the possibility that the GCs causing marginal skewness in the measured velocity distribution are associated with the stream of stars that were found crossing NGC 4365 (North East to South West) from analysis of wide-field, very deep $B$ filter imaging (\citealt{Bo12}; Mihos et al.\ 2012, in prep). The stream seems to be an indication of an ongoing merger with a small galaxy. The distinct velocities of these GCs may indicate that they are not associated NGC 4365's GC system and the recently discovered stream is a likely candidate. Fig. \ref{fig:sigma} shows the GC line-of-sight velocities plotted against galactocentric radius. To define which GC velocities are outliers we used the GC velocities $>1243$ km s$^{-1}$ (where the velocity distribution is not affected by skewness) to calculate the standard deviation of the sample, $\sigma$. The calculation of $\sigma$ was done in radial bins of 29 GCs each (excluding borderline velocity measurements). The calculated $1\sigma$, $2\sigma$ and $3\sigma$ envelopes are plotted above and below the systemic velocity and show clearly that there is a larger number of GCs with low velocities outside the $2\sigma$ envelope (13) than with high velocities (3). 

\begin{figure}\centering
 \includegraphics[width=0.49\textwidth]{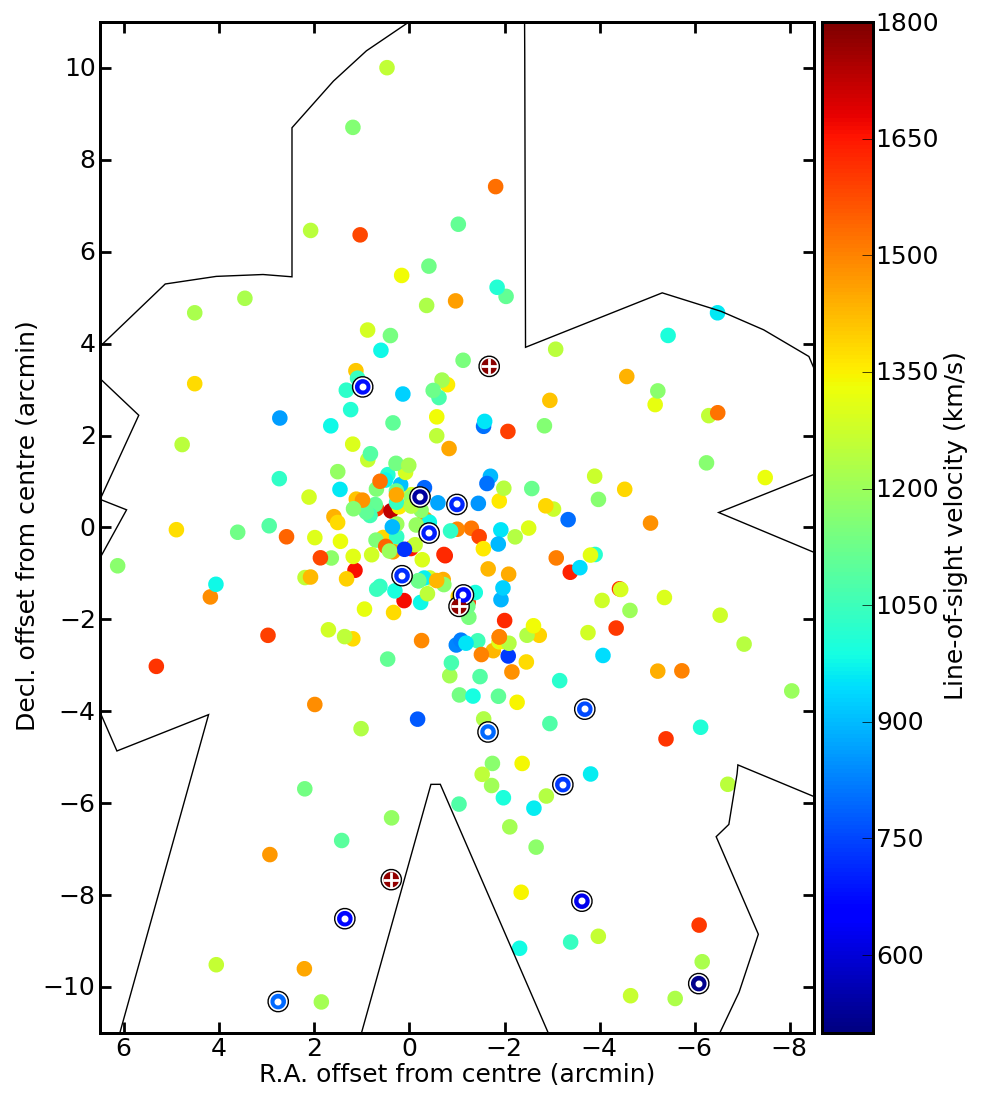}
 \caption{Spatial distribution of GCs showing individual radial velocities. The colour of individual points correspond to their radial velocity and the scale runs from 500 (dark blue) to 1800 km s$^{-1}$ (dark red). GCs with radial velocities more than $2\sigma$ from the host galaxy are marked with plusses (positive) and dots (negative).} 
 \label{fig:spatial}
\end{figure}

When we plot these velocity outliers on a map of GC surface density with colour and galactocentric radius there is no strong indication that the low velocity outliers follow a separate distribution to the colour-radius distribution of NGC 4365 GCs (see Fig. \ref{fig:dens}). The density map shows the distribution of photometrically identified GCs overplotted with the kinematic sample of GCs (those with reliable radial velocities). The blue and red subpopulations, as well as an overdensity of green GCs in the central regions, are visible in the density plot and the GCs with radial velocity measurements follow the colour-radius distribution well. The low velocity outliers tend to have red or green colours in the inner regions and blue colours in the outer regions of the galaxy, which is consistent with the general distribution of NGC 4365 GCs in colour-radius space.

In Fig. \ref{fig:spatial} the positions of the low velocity GCs are plotted on a map of the spatial distribution of the GCs with line-of-sight velocities. All but two of the low velocity GCs form a line running NE to SW across NGC 4365. This spatial structure of low velocity GCs is aligned with the recently discovered tidal stream (\citealt{Bo12}; Mihos et al.\ 2012, in prep) and would seem to indicate that the majority of the low velocity GCs are associated with the stream.

In summary, the evidence does not conclusively show that the low velocity GCs are associated with the stellar stream. Their distribution in colour-radius space is not distinguishable from the overall distribution of NGC 4365 GCs but their spatial distribution does suggest that they are different from the bulk of NGC 4365 GCs. We exclude all GCs with line-of-sight velocities that lie outside the $2\sigma$ envelopes and GCs with borderline velocity measurements from further kinematic analysis. This conservative sample ensures that the kinematic results are unaffected by unreliable GC velocity measurements and that the sample is not skewed to low velocities. We note here that including the low velocity outliers makes no significant difference to subsequent kinematic results, except to increase the uncertainties in our fits.

\section{Kinematics of the GC subpopulations}

\subsection{Kinematic model description}

To investigate the detailed kinematic properties of the three GC subpopulations we fit an inclined disk model to the GC system of NGC 4365, using the individual GCs as tracers. This inclined disk model is able to fit for rotation velocity and the position angle of that rotation with our sparse and non-uniform sampling of tracer particles. 
To fit this model we minimize \citep[as in][]{Fos11};

\begin{equation}
\Lambda = \sum_{i=1}^{N}\left[\frac{(V_{obs,i}-V_{mod,i})^2}{\sigma^{2}+(\Delta V_{obs,i})^2}+\ln(\sigma^{2}+(\Delta V_{obs,i})^2)\right]
\end{equation}

\begin{equation}
V_{mod,i} = V_{sys}\pm\frac{V_{rot}}{\sqrt{1+\left(\frac{\tan(\mathrm{PA}_{i}-\mathrm{PA}_{kin})}{q_{kin}}\right)^2}}
\end{equation}

where $V_{obs,i}$ is the observed line-of-sight velocity, $\Delta V_{obs,i}$ is the uncertainty on the observed velocity, $\sigma$ is the fitted velocity dispersion of the sample, $V_{mod,i}$ is the modelled velocity given by equation (4), $V_{sys}$ is the recession velocity of NGC 4365, $V_{rot}$ is the rotation velocity fitted to the sample, PA$_{i}$ is the position angle of the $i^{\mathrm{th}}$ GC, PA$_{kin}$ is the kinematic position angle fitted to the sample and $q_{kin}$ is the kinematic axis ratio ($b/a$) fixed for the sample. In contrast to \citet{Fos11} we do not use a minimisation algorithm to determine the minimum value of $\Lambda$ but evaluate the function for all reasonable values of $V_{rot}$, $\sigma$ and PA$_{kin}$. This brute force method ensures that we have found the global minimum for the function and therefore the best fitting values.

We use the bootstrap Monte Carlo method to obtain an estimate of the uncertainty of the model fit. A random sample of GCs is chosen from the original GC sample (equal in number to the original sample but allowing the data from a given GC to be used more than once) and then $\Lambda$ is minimized for that randomised sample. This is repeated 1000 times and the uncertainties are determined by calculating the central 68 per cent confidence intervals of the distribution ($1\sigma$) for each of $V_{rot}$, $\sigma$ and PA$_{kin}$. To use a brute force minimisation method for each iteration would be prohibitively time consuming and therefore we use a Powell minimisation algorithm (from \textsc{SCIPY.OPTIMIZE} in the \textsc{PYTHON} framework) with initial estimates set to the solution obtained from the kinematic best fitting values. The Powell minimisation algorithm is marginally sensitive to the initial conditions but we found no significant difference in the uncertainties obtained using this method and those found for a test case where the brute force method was used instead.

In general the $1\sigma$ uncertainties are not symmetric about the best fitting value because the distribution obtained from the bootstrap algorithm is not symmetric about the peak value. We calculate a central 68 per cent confidence interval rather than a symmetric one. The peak value of the distribution aligns well with the best fitting value in most cases. We do not artificially align the peak value of the distribution with the best-fit value and consequently it is possible for the best fitting value to fall outside the $1\sigma$ range of the distribution. This is a clear indication that for such a case the model is inappropriate for the data.

\subsection{Kinematics as a function of colour}

\begin{figure}\centering
 \includegraphics[width=0.49\textwidth]{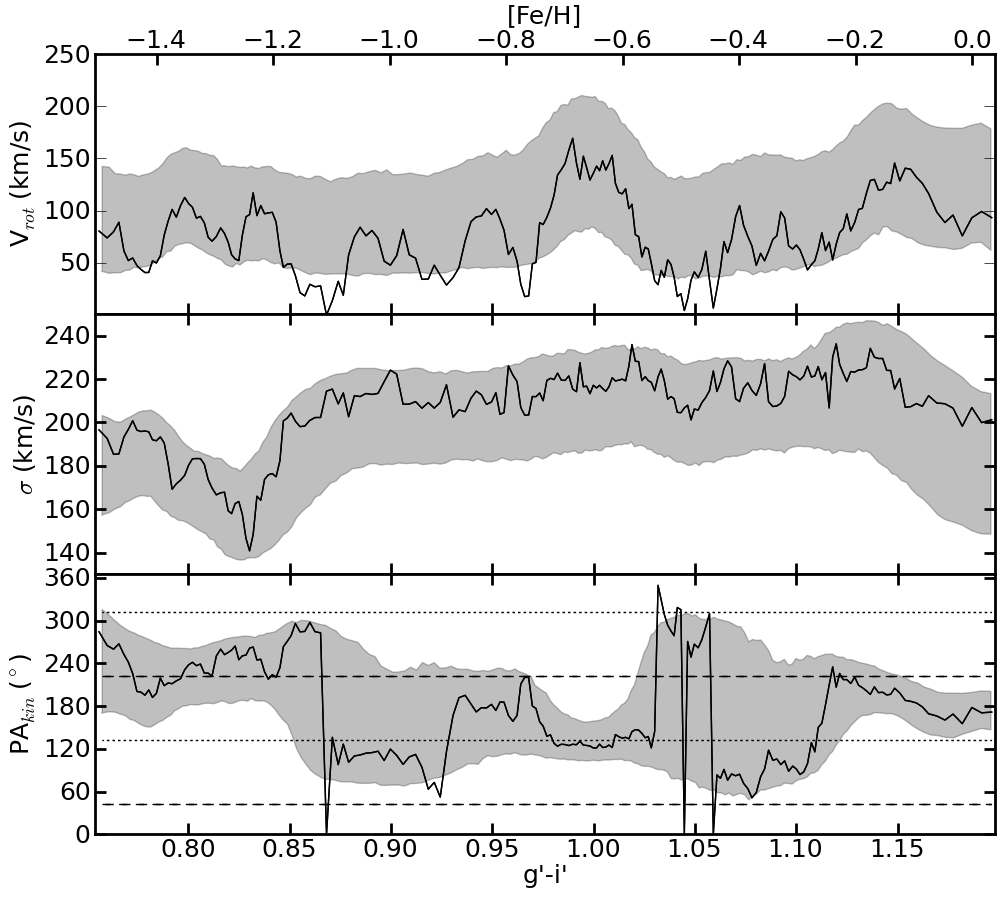}
 \caption{Kinematics for NGC 4365 GCs as a function of $g'-i'$ colour.  Kinematics are fitted for a rolling sample of 30 GCs and are plotted at the mean colour of the sample. The top panel shows the fitted rotation velocity ($V_{rot}$), the middle panel shows the fitted velocity dispersion ($\sigma$) and the bottom shows the fitted kinematic position angle (PA$_{kin}$). Solid lines show the kinematic fit to the data and the grey filled area shows the $1 \sigma$ uncertainties (obtained from 1000 bootstrap trials). On the bottom panel the photometric major axis of NGC 4365 \citep{Blom12} is plotted with dashed lines and the minor axis is plotted with dotted lines. 
 The bottom panel shows that the very blue, green and very red GCs all have constrained position angles of rotation and that the blue and red subpopulations rotate in a different direction to the green subpopulation. Where there is significant contamination between subpopulations in colour the rotation is unconstrained.}
 \label{fig:colkin}
\end{figure}

We use our kinematic model to analyse the kinematics of NGC 4365's GCs as a function of colour. The GCs with velocities less than $2\sigma$ away from the galaxy systemic velocity and secure velocity measurements are used to form a rolling sample of 30 GCs as follows. The 30 GCs with bluest colours form the first subsample. To form the next subsample the bluest GC is discarded while the GC with the bluest colour that was not included in the first subsample (i.e.\ the next reddest) is added and this process is repeated until all GCs have been included in a subsample. We chose the sample size of 30 as a best compromise between samples large enough to robustly constrain the kinematics and samples small enough to give adequate colour resolution. The kinematic fitting procedure is carried out on each subsample and the results are shown in Fig. \ref{fig:colkin}. In this case we fix the kinematic axis ratio to one ($q_{kin}=1$) as the photometric axis ratio for the galaxy light as well as blue, green and red GC subpopulations are all different \citep[$q_{blue}=0.56$, $q_{green}=0.70$, $q_{red}=0.97$ and $q_{gal}=0.75$ from][]{Blom12}. This assumption of $q_{kin}=1$ is equivalent to assuming a circular rotation disk and is chosen as the simplest case scenario.

\begin{table*}\centering
\caption{Summary of kinematic fits to various divisions of the three GC subpopulations. We tabulate the rotation velocity ($V_{rot}$), velocity dispersion ($\sigma$) and kinematic position angle (PA$_{kin}$) for each subpopulation in the medium probability (probability denoted with $P_{Colour}$) and best colour cut splits. Numbers of GCs in each subpopulation representative sample are given in brackets after the definition parameters. The major axis of NGC 4365's stellar light is $42^\circ$ or $222^\circ$.}
\begin{tabular}{cccccccccc}\hline
Split & $V_{rot}$ & $\sigma$ & PA$_{kin}$ & $V_{rot}$ & $\sigma$ & PA$_{kin}$ & $V_{rot}$ & $\sigma$ & PA$_{kin}$ \\
Definition  & (km s$^{-1}$) & (km s$^{-1}$) & ($^\circ$) & (km s$^{-1}$) & (km s$^{-1}$) & ($^\circ$) & (km s$^{-1}$) & (km s$^{-1}$) & ($^\circ$) \\ \hline
medium probability & \multicolumn{3}{c}{$P_{Blue}>0.8$ (53)} & \multicolumn{3}{c}{$P_{Green}>0.8$ (53)} & \multicolumn{3}{c}{$P_{Red}>0.8$ (65)} \\ \noalign{\smallskip}
& $49\pm^{48}_{18}$ & $179\pm^{10}_{17}$ & $254\pm^{38}_{54}$ & $83\pm^{57}_{29}$ &  $229\pm^{13}_{22}$ & $145\pm^{33}_{31}$ & $39\pm^{52}_{11}$ &  $218\pm^{14}_{22}$ & $174\pm^{62}_{53}$ \\ \hline
colour cut & \multicolumn{3}{c}{$g'-i'<0.85$ (61)} & \multicolumn{3}{c}{$0.9<g'-i'<1.05$ (77)} & \multicolumn{3}{c}{$g'-i'>1.1$ (56)} \\ \noalign{\smallskip}
& $67\pm^{44}_{27}$ & $178\pm^{10}_{16}$ & $269\pm^{28}_{34}$ & $81\pm^{39}_{24}$ &  $217\pm^{10}_{15}$ & $144\pm^{27}_{23}$ & $94\pm^{48}_{34}$ &  $202\pm^{15}_{26}$ & $194\pm^{23}_{24}$ \\ \hline
\end{tabular}
\label{tab:kin}
\end{table*}

Fig. \ref{fig:colkin} shows that the GCs of NGC 4365 have significantly different kinematic features, which depend on GC colour. The changes in position angle seen in Fig. \ref{fig:colkin} show this very clearly, e.g.\ blue GCs 
rotate with PA$_{kin}$ that is consistent with the photometric major axis, green GCs rotate with PA$_{kin}$ consistent with the photometric minor axis and red GCs with rotate with PA$_{kin}$ that is between the photometric major and minor axes of NGC 4365. At the colours where we expect significant overlap between subpopulations the rotation cannot be constrained. 
Here the values fitted for the PA$_{kin}$ become unstable while the values fitted for $V_{rot}$ drop to zero and become inconsistent with the errors calculated. Where it is not possible to constrain the position angle of the rotation, the amplitude of rotation is also unconstrained. 

\citet{Fos11} noted that $V_{rot}$ is artificially increased when the PA$_{kin}$ is allowed to vary. \citet{St11} describe this artificial increase as a rotation bias and note that this bias is important when the amplitude of rotation is low, e.g.\
\begin{equation}
\frac{V_{rot}}{\sigma} \leqslant 0.55\sqrt{\frac{20}{N}}
\label{eq:vbias}
\end{equation}
Here $N$ denotes the number of GCs in the sample. When we apply this criterion to the kinematic fit to NGC 4365's GCs as a function of colour the colour ranges where the PA$_{kin}$ is well constrained ($g'-i' \sim 0.79-0.84$, $0.97-1.02$ and $1.12-1.17$) the rotation parameter ($V_{rot}/\sigma$) is well above the criterion value for this sample size ($N=30$). We also see that the rotation parameter fails the criterion (i.e.\ is affected by rotation bias) in areas where the PA$_{kin}$ is poorly constrained (e.g.\ $0.87<g'-i'<0.95$) even when the fitted $V_{rot}$ is measured to be significantly greater than zero.

In the blue, green and red colour ranges where we do constrain the kinematics the blue subpopulation $V_{rot}$ and $\sigma$ ranges are $50 - 140$ and $140 - 160$ km s$^{-1}$ respectively while the green subpopulation $V_{rot}$ and $\sigma$ ranges are $80 - 200$ and $180 - 220$ km s$^{-1}$ respectively. The rotation of the red subpopulation is range is $V_{rot} \sim 80 - 180$ km s$^{-1}$ while its velocity dispersion decreases with increasingly red GC colour, $\sigma \sim 220$ to $\sim 200$ km s$^{-1}$.

This observation of distinct kinematic features is a confirmation that the three GC subpopulations of NGC 4365 seen in colour and other photometric properties \citep{La03,Br05,Blom12} are indeed separate. They are likely to have formed via different mechanisms or at different epochs. To better constrain the position angle (PA$_{kin}$) and derive clearer values for the amplitude of rotation ($V_{rot}$) and dispersion ($\sigma$), we need to divide the GCs into three uncontaminated representative samples of the three GC subpopulations.

\subsection{Dividing the sample into three subpopulations}

\begin{figure*}\centering
 \includegraphics[width=0.98\textwidth]{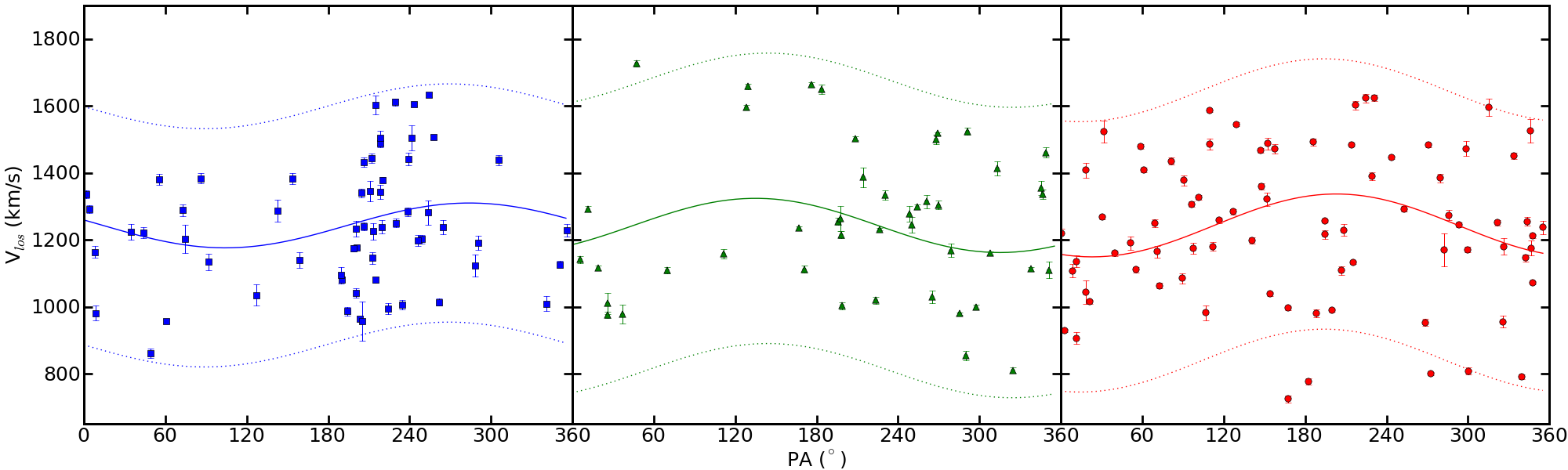}
 \caption{Observed line-of-sight velocity is plotted against position angle for the blue (left), green (centre) and red (right) GC subpopulations. We show the subpopulations as defined by the best colour cut split (see Table \ref{tab:kin}). The fitted rotation velocity is plotted with a solid sinusoid, and an envelope of twice the fitted velocity dispersion is shown with dotted lines for each subpopulation. The position angle (PA$_{kin}$) of rotation is $269^\circ$, $144^\circ$ and $194^\circ$ for the blue, green and red subpopulations respectively. The photometric major axis of the galaxy is $42/222^\circ$ and the overdensity of blue GCs at $\sim210^\circ$ is due to the strong elongation of the blue GC subpopulation along the major axis of the galaxy.}
 \label{fig:velpa2}
\end{figure*}

\citet{Blom12} divided NGC 4365's GCs into three subpopulations according to a probability of belonging to that subpopulation. They assigned each GC a probability of belonging to the blue, green and red subpopulations ($P_{Blue}+P_{Green}+P_{Red}=1$) based on its colour and galactocentric radius. ``The process used to determine the blue-green-red probability of a GC starts by determining the relative number of GCs in each subpopulation at the galactocentric radius of the GC. This is done by comparing the values of the radial surface density profile for each subpopulation. The relative numbers of GCs in each population are used to scale the normalisations of the three Gaussian distributions and comparing the relative values of all three Gaussian distributions at the colour of the GC [they] calculate the probability of an object being blue, green, or red." \citep{Blom12}  Their high probability subsamples ($P>0.95$) do not include any green GCs and therefore we define medium probability samples with $P>0.80$ to represent the three subpopulations. Of GCs with radial velocities, 53 blue, 53 green and 65 red GCs form part of the medium probability samples.

The kinematic fits to the medium probability fits are shown in Table \ref{tab:kin}. The blue and green sample fits agree with the kinematic properties found in Fig. \ref{fig:colkin} but the fit to the red sample is not well constrained (PA$_{kin}$ varies by 115$^\circ$).The blue medium probability sample contains GCs with colours bluer than $g'-i'\sim0.85$ and the green sample contains GCs with colours between $g'-i'\sim0.85$ and $g'-i'\sim1.05$. These samples contain relatively few GCs but do not overlap and are mostly free of contamination from the other subpopulation. The red medium probability sample contains more GCs but also contains GCs with colours as `blue' as $g'-i'\sim1.0$ and is likely affected by contamination from green GCs. We expect some contamination particularly in the red sample because the medium probability red GCs extend almost as far as the peak colour of the green subpopulation \citep[$g'-i'=0.98$,][]{Blom12}. The probability that these GCs belong to the green subpopulation is close to 20 percent and not trivial.

We also experimented with splitting the GCs into three subsamples at different $g'-i'$ colours, detailed in Appendix A. To determine the most accurate kinematics for the three colour subpopulations of NGC 4365 we need to include as many GCs as possible in each sample while minimising the number of GCs from one subpopulation contaminating the sample for another subpopulation. The best kinematic results are determined by the fit which has the minimum value of $\Lambda/ndf$ (see equation 3) and the most constrained PA$_{kin}$. Salient results are shown in Table \ref{tab:kin} and details discussed in Appendix A.  The best kinematic fits are found for the 61 blue GCs with $g'-i'<0.85$, 77 green GCs with $0.9<g'-i'<1.05$ and 56 red GCs with $g'-i'>1.1$. The results for the medium probability and best colour cut samples are consistent within the uncertainties, but the best colour cut sample contains larger numbers of blue and green GCs and a relatively uncontaminated sample of red GCs. The red sample in this case ends close to the expected peak in the red subpopulation ($g'-i'=1.13$). Henceforth we use the best colour cut sample for the subpopulation representative samples.

The kinematics are fitted twice for each colour split definition, once with the axis ratio fixed to the photometric value for each GC subpopulation ($q_{kin}=q_{photom}$) and once with the axis ratio fixed to unity ($q_{kin}=1$) as before. We do not have enough GCs in each sample to fit for $q_{kin}$ and it is not clear that the kinematic axis ratio would necessarily be equal to the photometric axis ratio, especially since Fig. \ref{fig:colkin} leads us to expect rotation in different directions to the photometric position angle. We do not find any significant differences in the fitted kinematics between the two axis ratio cases and conclude that this kinematic fitting method is not very sensitive to the kinematic axis ratio (see Appendix A). All further kinematic fits are done in the case where $q_{kin}=1$ as this is the more general case.

We compare the rotation of these subpopulation representative samples (henceforth referred to as the blue, green and red subpopulations) with each other and with the kinematics of the central starlight (to $\sim0.5$ arcmin) as measured by \citet{Da01} and \citet{Kr11}. They found that the KDC rotates with a PA$_{kin}$ of $38\pm5^\circ$ aligned with the photometric major axis of the galaxy ($42^\circ$) and that the bulk of the starlight rotates with a PA$_{kin}$ of $145\pm6.5^{\circ}$, very close to the minor axis of the galaxy ($132^\circ$). Using the Stellar Kinematics from Multiple Slits (SKiMS) technique \citep[see also][]{Pr09,Fos11}, Arnold et al.\ (2012, in prep) find that the minor axis rotation of the starlight continues out to $\sim4.5$ arcmin. We find that the blue subpopulation rotates with a PA$_{kin}$ between $235^\circ$ and $297^\circ$, not consistent with either the major ($222^\circ$) or the minor ($312^\circ$) axes of NGC 4365. The green subpopulation rotates with a PA$_{kin}$ of $144\pm^{27\circ}_{23}$, consistent with the minor axis of the galaxy and the rotation of the bulk of the starlight. The red subpopulation rotates with a PA$_{kin}$ between $170^\circ$ and $217^\circ$, almost consistent with the major axis but in the \textit{opposite direction} to the stars of the KDC. The kinematic properties that we have found by using the best representative samples for each GC subpopulation agree with the kinematics seen in Fig. \ref{fig:colkin} where GCs kinematics are plotted as a function of colour.

Fig. \ref{fig:velpa} shows the combined rotation velocity and velocity dispersion. These are plotted separately for each subpopulation and the best-fit models show good agreement with the observed radial velocities of the subpopulations. The non-uniform azimuthal distribution of GCs, seen particularly in the blue subpopulation, but also in the green subpopulation, is partly due to the elliptical spatial distribution of these subpopulations \citep[axis ratio $q_{blue}=0.56$ and $q_{green}=0.70$, see][]{Blom12} and partly due to the asymmetric spatial coverage of the spectral observations. We do not expect this azimuthal bias to significantly influence the kinematic fits.

\subsection{Radial kinematics for three subpopulations}

\begin{figure}\centering
 \includegraphics[width=0.49\textwidth]{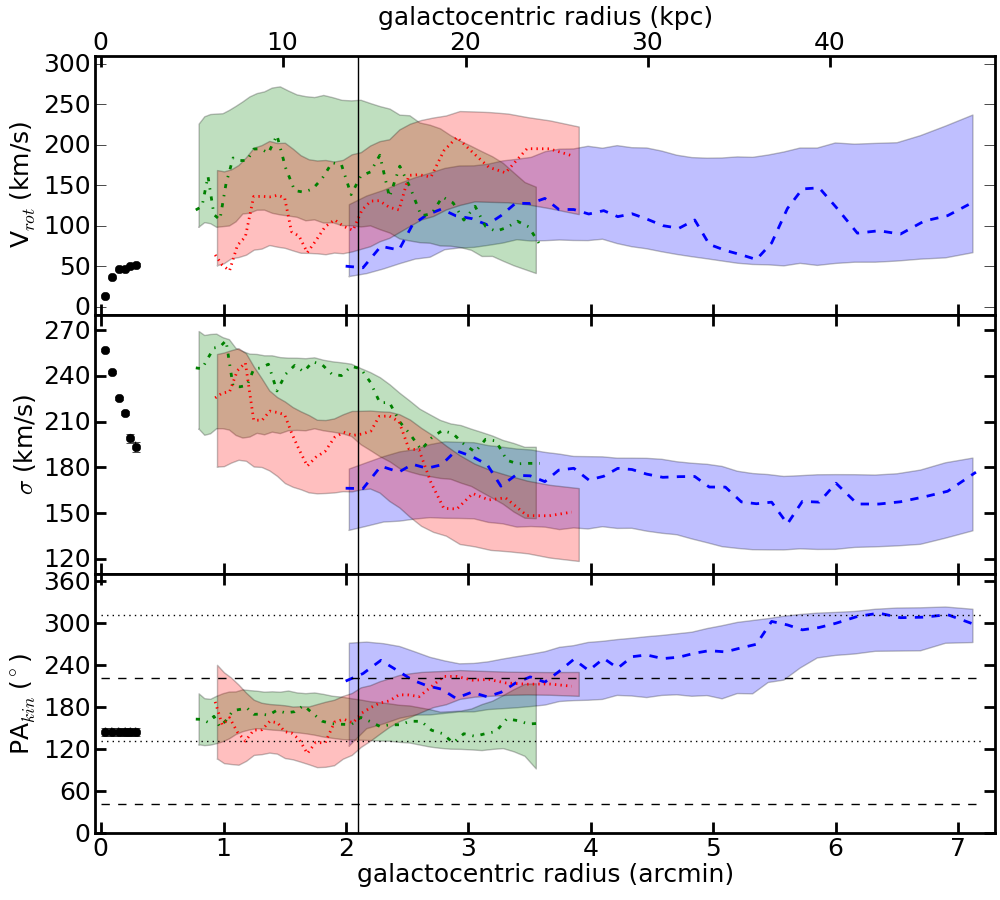}
 \caption{Kinematics as a function of radius for NGC 4365's three GC subpopulations. GC subpopulations are defined using the best colour cut in Table \ref{tab:kin}. Kinematics are fitted for a rolling sample of 22 GCs and are plotted at the mean galactocentric radius of the sample. The top panel shows the fitted rotation velocity ($V_{rot}$), the middle panel shows the fitted velocity dispersion ($\sigma$) and the bottom panel shows the fitted kinematics position angle (PA$_{kin}$). The blue subpopulation (all GCs bluer than $g'-i'=0.85$) is plotted with blue dashed lines, the green subpopulation (GCs between $g'-i'=0.9$ and $1.05$ with galactocentric radius $<5$ arcmin) is plotted with green dash-dotted lines and the red subpopulation (all GCs redder than $g'-i'=1.1$) is plotted with red dotted lines. The lines show the kinematic fit to the data and the filled areas show the $1 \sigma$ uncertainties (obtained from 1000 bootstrap trials). The black points show stellar kinematics extracted along the minor axis of NGC 4365 from the SAURON data \citep{dZ02}. The effective radius of the galaxy starlight is plotted with a vertical line at 2.1 arcmin \citep{Blom12}. On the bottom panel the photometric major axis of NGC 4365 is plotted with dashed lines and the minor axis is plotted with dotted lines. The variation of the PA$_{phot}$ is less than $3^\circ$ and not visible on this scale. The kinematic fits to the three subpopulations show clearly different features.}
 \label{fig:radkin}
\end{figure}

We use the best colour cut samples to analyse the variation of kinematics with galactocentric radius for each subpopulation. The samples contain 61 and 56 GCs for the blue and red subpopulations respectively, but only 61 of 77 GCs for the green subpopulation as we restrict the green sample to include only GCs with galactocentric radii smaller than 5 arcmin. We impose this galactocentric radius restriction on the green subpopulation because beyond this radius it contributes less than ten per cent of the total GC population. The GC subpopulation kinematics are fitted as a function of galactocentric radius in rolling bins of 22 GCs and the results of the kinematic fits are shown in Fig. \ref{fig:radkin}. The bin size of 22 GCs was chosen to maximize the radial range of the kinematic analysis. This is significantly smaller than the bin size used for Fig. \ref{fig:colkin} because inter subpopulation contamination has been reduced. We fit the kinematics in the case where the PA$_{kin}$ is allowed to vary with galactocentric radius. 

The blue subpopulation shows significant rotation at all radii. There is an indication that the PA$_{kin}$ of rotation twists from $\sim 200^\circ$ to $\sim 310^\circ$ between 3 and 7 arcmin (20 and 47 kpc). 
The rotation velocity is consistent with being flat and $V_{rot}\sim 100$ km s$^{-1}$ over the entire radial range of the blue subpopulation. The velocity dispersion of the blue subpopulation is also consistent with being flat and $\sigma \sim 200$ km s$^{-1}$. 
The radial kinematic profiles are consistent with the values found for the blue GCs as seen in Fig. \ref{fig:colkin} as well as the fitted values for the entire blue subpopulation sample (Table \ref{tab:kin}). The kinematics of the blue GC system varies more significantly with galactocentric radius than GC colour and by removing most of the contaminating GCs coming from the green subpopulation as well as dividing the blue GC by radius, we achieve tighter constraints on the subpopulation kinematics. The twist in the PA$_{kin}$ could suggest that the radial outskirts of the blue subpopulation have been affected by accretion of galaxies and associated GCs, or a tidal interaction. This kinematic twist could also be a signature of triaxiality in the blue GC subpopulation.

The green subpopulation rotates relatively fast at small radii ($V_{rot}\sim200$ km s$^{-1}$) but the rotation decreases quickly with increasing galactocentric radius to $\sim75$ km s$^{-1}$ at 3.5 arcmin. The radially fitted PA$_{kin}$ is consistent with being flat ($150\pm30^\circ$) and consistent with the PA$_{kin}$ fitted to the entire subpopulation ($144\pm^{27\circ}_{23}$). 
The velocity dispersion of the green subpopulation also decreases with galactocentric radius from $260$ to $180$ km s$^{-1}$.

The red subpopulation rotates fastest between 2 and 4 arcmin, with PA$_{kin}\sim 222^\circ$. The rotation velocity amplitude here is $V_{rot}\sim 180$ km s$^{-1}$.
The fitted PA$_{kin}$ within 2 arcmin is not as well constrained as that fitted between 2 and 4 arcmin, and within 2 arcmin the rotation parameter fails the criterion defined in Equation \ref{eq:vbias}. The rotation of the red GC subpopulation at small galactocentric radii is consistent with zero rotation and the value of $\sim75$ km s$^{-1}$ fitted here is due to rotation bias caused by the free variation of PA$_{kin}$. The velocity dispersion of the red GCs decreases with radius from $\sigma \sim 220$ to $\sim 150$ km s$^{-1}$. The red subpopulation is dispersion dominated in the central regions and rotation dominated in the outer regions of the galaxy.

\begin{figure}\centering
 \includegraphics[width=0.49\textwidth]{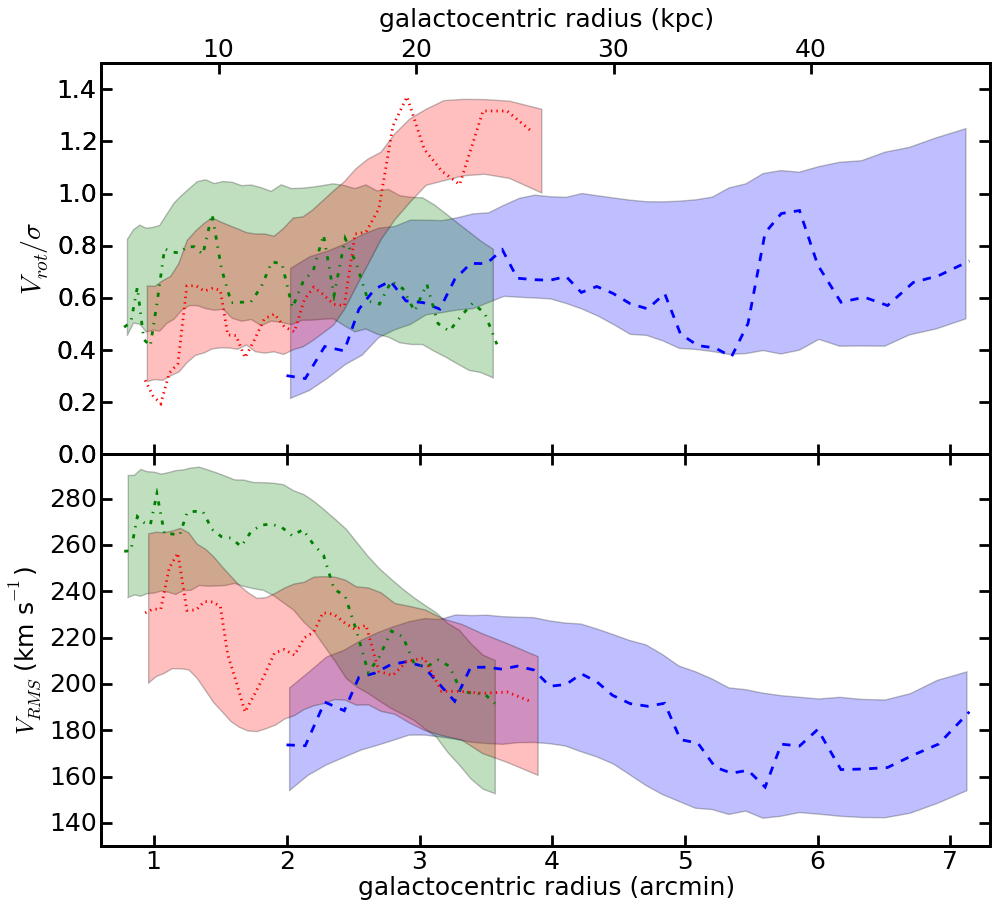}
 \caption{Kinematic parameters of GCs as a function of galactocentric radius. Rotation parameter $V_{rot}/\sigma$ (top panel) as well as the root mean squared velocity $V_{RMS}$ (bottom panel) for GC subpopulations potted against galactocentric radius. Linestyles are the same as those in Fig. \ref{fig:radkin}. This figure shows the case where the position angle is allowed to vary. The rotation bias is small everywhere except inside a galactocentric radius of 2 arcmin for the red subpopulation. Here the plotted value for $V_{rot}/\sigma$ is significantly overestimated but because there is no fitting involved in obtaining the $V_{RMS}$ this value is not affected.}
 \label{fig:vsig}
\end{figure}

\begin{table*}\centering
\caption{Summary of the NGC 4365 galaxy system properties. Values marked with $^\dagger$ are taken from \citet{Blom12}, $^\diamond$ from \citet{Kr11}, $^\star$ from \citet{VCS6} and $^\bullet$ from \citet{Da01}, while all other values are derived in this work. Notes: $^1$The position angles for the blue and green subpopulations were qualitatively similar to the galaxy light and were therefore assumed equal to it. $^2$This effective radius measurement is not obtained directly from a S\'{e}rsic profile fit to the green subpopulation but by comparing the power law slope of the green and red subpopulations. $^3$The $g'-z'$ value was converted to a $g'-i'$ colour value. $^4$These are maximum velocities observed in a limited field-of-view not mean or median velocities. $^5$This is the largest radial extent of the KDC and not its effective radius.}
\begin{tabular}{cccccccc}\hline\noalign{\smallskip}
& $g'-i'$ (mag) & PA$_{photom}$ $(^{\circ})$ & PA$_{kin}$ $(^{\circ})$ & $V_{rot}$ (km s$^{-1}$) & $\sigma$ (km s$^{-1}$) & $q_{phot}$ & $R_e$ (arcmin) \\ \hline\noalign{\smallskip}
Blue GCs & 0.80 $^\dagger$ & $\sim 42/222$ $^{\dagger,1}$ & $269\pm^{28}_{34}$ & $67\pm^{44}_{27}$ & $178\pm^{10}_{16}$ & $0.56\pm0.08$ $^\dagger$ & $7.30\pm0.68$ $^\dagger$ \\\noalign{\smallskip}
Green GCs & 0.97 $^\dagger$ & $\sim 42/222$ $^{\dagger,1}$ & $144\pm^{27}_{23}$ & $81\pm^{39}_{24}$ & $217\pm^{10}_{15}$ & $0.70\pm0.08$ $^\dagger$ & $\sim 2.97$ $^{\dagger,2}$ \\\noalign{\smallskip}
Red GCs & 1.13 $^\dagger$ & undefined & $194\pm^{23}_{24}$ & $94\pm^{48}_{34}$ & $202\pm^{15}_{26}$ & $0.97\pm0.08$ $^\dagger$ & $3.17\pm0.19$ $^\dagger$ \\\noalign{\smallskip}
Stars & 1.15 $^\dagger$ & $42/222\pm3$ $^\dagger$ & $145\pm6.5$ $^\diamond$ & 60.9 $^{\diamond,4}$  & $260\pm10$ $^{\bullet,4}$ & $0.75\pm0.03$ $^\dagger$ & $2.1\pm0.11$ $^\dagger$ \\\noalign{\smallskip}
KDC & 1.40 $^{\star,3}$ & $40.9\pm2.1$ $^\diamond$ & $38\pm5$ $^\bullet$ & $63\pm8$ $^{\bullet,4}$ & $262\pm3$ $^{\bullet,4}$ & $0.75\pm0.08$ $^\dagger$ & $0.12\pm0.02$ $^{\bullet,5}$ \\ \hline\noalign{\smallskip}
\end{tabular}
\label{tab:summ}
\end{table*}

Plotting the rotation parameter ($V_{rot}/\sigma$) in Fig. \ref{fig:vsig} we see that the red subpopulation shows very different properties to the blue and green subpopulations. This ratio is one measure of the diskyness of the system, where $V_{rot}/\sigma>1$ can indicate a disky system. It is only the red GC subpopulation that shows rotationally dominated characteristics, and only at galactocentric radii beyond 2.5 arcmin, where it rotates with a PA$_{kin}$ of $\sim222^\circ$ (photometric major axis of NGC 4365). 

The root mean squared velocity $\left(V_{RMS}\sim\sqrt{\frac{\sum{(V_{los}-V_{sys})^2}}{N}}\right)$ shows variation with radius for all three subpopulations but changes most significantly in the green subpopulation. The green GC subpopulation $V_{RMS}$ decreases from $\sim 270$ to $\sim 190$ km s$^{-1}$ over 1.5 arcmin in galactocentric radius. The sharply declining $V_{RMS}$ seen in the case of the green subpopulation could be explained by a truncation in their density distribution or a bias towards radial orbits relative to the other GC subpopulations.


\section{Discussion}

Since most giant elliptical galaxies contain only two metallicity subpopulations of GCs it is likely that one of NGC 4365's three GC subpopulations is the result of an additional and unusual formation/accretion process. Determining which of the GC subpopulations is additional may indicate which mechanism caused it and could be a key to uncovering the formation mechanisms of GC subpopulations in ordinary elliptical galaxies. We summarise the kinematic and photometric properties of the three GC subpopulations as well as the two main components of the galaxy starlight (the KDC and the bulk of the stars) in Table \ref{tab:summ} and use these properties to discuss possible formation scenarios.

\begin{figure}\centering
 \includegraphics[width=0.49\textwidth]{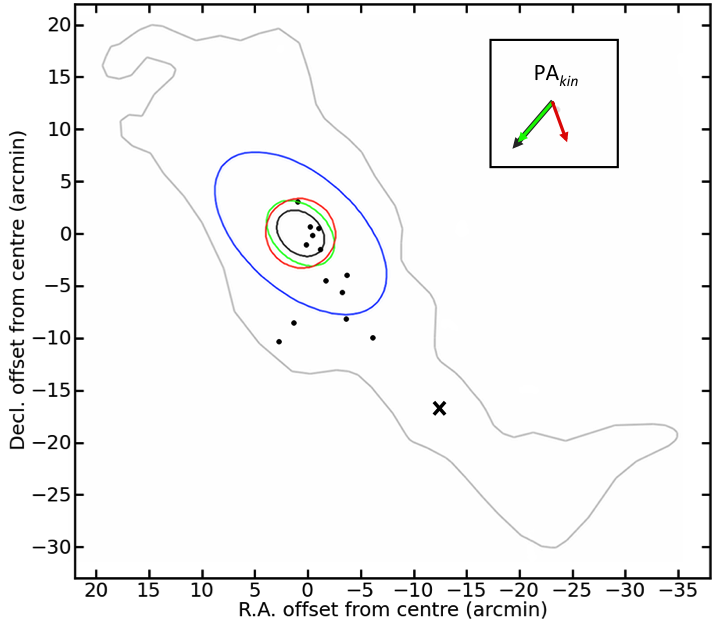}
 \caption{Schematic representation of the NGC 4365 stellar light and GC system. The diagram is oriented with North up and East left. Positions of low velocity GC outliers are shown with black dots. Ellipses show the 2$R_e$ extent of the galaxy light (black) and three GC subpopulations in blue, green and red respectively. The shape and relative size of the stellar light stream \citep[reported in][]{Bo12} is shown with a grey irregular outline. The rotation direction of the galaxy light as well, as the green and red GC subpopulations, are shown in the top right of the diagram. The rotation direction of the blue GCs is not shown because there is a significant twist with galactocentric radius. A cross marks the position of NGC 4342. The low velocity GCs are generally aligned with the stellar stream and have a similar velocity to NGC 4342.}
 \label{fig:schm}
\end{figure}

We find three distinct kinematic signatures for the three GC subpopulations. The blue subpopulation rotates at all observable radii and its kinematic position angle twists slowly with increasing galactocentric radius from  $200^\circ$ to $312^\circ$. Its rotation is unrelated to NGC 4365's stellar light. The green subpopulation rotates with the bulk of NGC 4365's stellar light along the photometric minor axis of the galaxy ($132^\circ$). The red subpopulation rotates only at large radii along the photometric major axis ($222^\circ$) of NGC 4365 but in the opposite direction to the KDC of the galaxy ($38^\circ$). 
The distinct kinematics of the red and green GC subpopulations, seen clearly in Fig. \ref{fig:schm}, provide evidence that the green GCs form a separate subpopulation and are not simply the metal poor tail of the red subpopulation.

The \citet{Blom12} photometric analysis found that the radial density distribution and colour of the red GC subpopulation most closely matched the stellar light of NGC 4365 and, while the axis ratio of the green subpopulation more closely matched that of the starlight, they concluded that there was evidence that the red subpopulation was related to the bulk of the galaxy starlight. The analysis here shows that the rotation position angle, and possibly also the rotation velocity amplitude, of the green GC subpopulation most closely match the extrapolated stellar kinematics. Thus it is not immediately clear which of the two is the additional subpopulation.

If the red GC subpopulation is additional it could have been created by an additional, later (still $>10$ Gyr ago) merger \citep{Ze93}. This might explain why the red subpopulation rotation is decoupled from the bulk of the starlight and the green subpopulation rotates with the bulk of the stars. The problem arises when comparing this red subpopulation to those of other galaxies of similar luminosity to NGC 4365, as it is not an outlier but consistent with the colour of red subpopulations of other galaxies of similar luminosity/mass \citep{VCS9}. 

If the green GC subpopulation has been recently donated from a medium sized galaxy that accreted on to NGC 4365 then the remnant of this accreted galaxy might still be visible in the form of a stellar stream across NGC 4365 as reported by \citet{Bo12}. \citet{Blom12} reported that the green subpopulation is centred at $g'-i' \sim 0.97$ and we can estimate the absolute magnitude range for the accreted galaxy by comparing the peak colour of the green subpopulation with the relationship between peak GC subpopulation colour and galaxy luminosity empirically determined by \citet{VCS9}. In this case the green subpopulation of NGC 4365 is the original metal-rich (red) GC subpopulation of a $-14<$ M$_{B} <-15$ dwarf galaxy and we expect to find primarily blue and green GCs in the stellar stream of NGC 4365. The total stellar light observed in the low surface brightness stream is $B \sim13.5$ mag \citep[][this a minimum brightness for the accreted galaxy]{Bo12} corresponding to M$_B \sim$ -18.3 at the distance of NGC 4365. There is a sizeable mismatch between the predicted absolute magnitude for the dwarf galaxy and the much brighter integrated stellar light from the stream, suggesting that the stellar stream is not the remnant of the galaxy associated with accreted green GCs. This explanation has several other problems. It is still to be explained why the accreted green subpopulation is so centrally concentrated while the stellar light of the accreted galaxy is still on the outskirts of NGC 4365. Also why does the green subpopulation rotate with the bulk of the stellar light of NGC 4365 (along the minor axis of the galaxy) rather than along the stream orientation, which is roughly aligned with the major axis of NGC 4365?

If the accretion event which brought the green GCs into NGC 4365 was not recent, the currently observable stellar stream may be unrelated to the green GCs. We can still investigate this claim based on the number of green GCs \citep[see a similar analysis by][]{Ro12}. \citet{Blom12} reported that the green subpopulation contributes 17 percent of the total GC population or $1100\pm20$ GCs to NGC 4365's GC system (N$_{Tot}=6450\pm110$). We can calculate the absolute magnitude range of the accreted galaxy using the observed relationship between GC specific frequency \citep[S$_N=N_{GC}10^{0.4(M_V+15)}$,][]{Ha81} and galaxy luminosity \citep{VCS15}. To be consistent with the S$_N$ range observed for GC systems in the literature, the accreted galaxy must be brighter than $M_V\sim-20.5$ ($M_B\sim-19.5$). The number of GCs in the green subpopulation far exceeds the expected number of GCs that would be associated with a dwarf galaxy. If the green subpopulation was donated by an accreted galaxy (at any stage in the history of NGC 4365) the mean colour of the green GCs suggests a very small galaxy whereas the number of green GCs suggests a large galaxy. 

If the green GCs were accreted from a group of dwarf galaxies it might account for the large number of GCs with colours that indicate a small host galaxy, but the group of dwarfs would need to contain many tens of members to account for the large number of inferred green GCs.


We also consider the possibility that both the red and green subpopulations were formed in-situ. In this case the green subpopulation would have formed unusually early in the dense core of the galaxy while the red GCs formed later when the core of the galaxy was slightly bigger and slightly more metal rich. This scenario might explain why the green subpopulation is more centrally concentrated than the red subpopulation and why the green GCs rotate with a significant fraction of the starlight. It does not explain why we do not detect a fraction of the stars rotating with the red GCs or what would have caused two epochs of early in-situ GC formation.

\citet{Bo12} studied the lenticular galaxy NGC 4342 \citep[$V_{sys}=751$ km s$^{-1}$,][]{redshifts} embedded in the stellar stream across NGC 4365 (see Fig. \ref{fig:schm}) and inferred a massive and extended dark matter halo around NGC 4342. The central velocity dispersions of NGC 4365 ($261\pm7$ km s$^{-1}$) and NGC 4342 ($244\pm11$ km s$^{-1}$) are consistent within errors \citep{vdisp} implying similar mass galaxies. It is possible that this very massive, if relatively faint, galaxy could have passed close by NGC 4365 and dragged some of NGC 4365's own stars out from the galaxy to form the stream. It is also possible that the stream stars have been stripped from NGC 4342 itself. These scenarios both agree with our observation of low velocity GCs (at $\sim 700$ km s$^{-1}$, similar to the $V_{sys}$ of NGC 4342) in NGC 4365 (Fig. \ref{fig:sigma}) that are aligned with the stellar stream (Figs \ref{fig:spatial} and \ref{fig:schm}). The low velocity GCs are consistent with the colour distribution of NGC 4365's underlying GC population (Fig. \ref{fig:dens}) but it is still unknown how their colours compare to those of NGC 4342. 

It is tempting to suggest that the close passage of NGC 4342 might have stretched NGC 4365 starlight and GC system into its current shape. If NGC 4365 was round or only mildly elliptical (with a PA$_{phot}$ close to the current PA$_{kin}$ of the bulk of stellar light) before NGC 4342 made its close pass we would expect to see remnants of the original shape in the central regions of NGC 4365. \citet{VCS6} find the galaxy ellipticity drops below 0.05 in the $g$ and $z$ bands while the $PA_{phot}$ twists to $\sim90^\circ$ at radii smaller than 0.3 arcsec. These photometric properties only appear at scales smaller than the KDC of NGC 4365 and are unlikely to relate to the bulk of NGC 4365. This scenario might explain the anomalous `rolling' starlight as an artefact of very recent perturbations to NGC 4365's photometric shape but would not explain the presence of NGC 4365's KDC . The KDC could be a remnant of an old merger, as its age is observed to be $>10$ Gyr and consistent with the galaxy starlight \citep{Da01} or the consequence of our particular viewing angle into an intrinsically triaxial system \citep{vdb91}.

Fig. \ref{fig:schm} summarises the photometric and kinematic information of the NGC 4365 system graphically. The blue GC subpopulation is the most radially extended GC subpopulation and is more elongated than the galaxy light. Its rotation shows a slow twist from alignment with the galaxy major axis in the inner parts to the galaxy minor axis in the outer parts. The amplitude of its rotation and velocity dispersion are constant with radius. The green subpopulation is the most centrally concentrated, roughly as elongated as the galaxy light and shows `rolling' kinematics like the galaxy light (PA$_{kin}\sim$ galaxy minor axis). Its rotation and velocity dispersion amplitude decrease with increasing radius. The red subpopulation is almost circular and rotates only in the outer parts, along the galaxy major axis (in the opposite direction to the KDC). The low velocity GCs are aligned with the stellar stream and contain GCs of all colour subpopulations. The kinematics of the galaxy and GC system are complex but it appears that the stellar stream is not connected to the presence of any one of the GC subpopulations.

\section{Conclusions}

NGC 4365 is a giant elliptical galaxy with rare stellar kinematic properties and also, very unusually, three globular cluster (GC) subpopulations. Recently, a stellar stream running across NGC 4365 and its nearby neighbour, NGC 4342, was discovered.

With high velocity resolution Keck/DEIMOS spectra we analyse the kinematics of a large number and spatially extended sample of GCs around NGC 4365. These data allow us to separate the GC system into its three colour subpopulations and analyse the kinematics of each subpopulation in detail. We find distinct rotation properties for each GC subpopulation. This indicates that the three GC subpopulations are distinct and that the formation history of NGC 4365 is complex. We conclude that it is unlikely that the third (green) GC subpopulation is related to the existence of the stellar stream and that it is also unlikely to have originated from an accreted galaxy.

We also find a further group of low velocity GCs (covering the full range in colour), which might be related to the stellar stream extending across NGC 4365 towards NGC 4342 ($\sim 35$ arcmin away at a similar systemic velocity to these GCs).  Future analysis of spectroscopic observations of GCs in the stream will further illuminate the complex formation history of NGC 4365. 

\section{Acknowledgements}
We thank the S.\ Larsen for constructive comments on the paper as well as A.\ Bogdan and C.\ Mihos for valuable discussion. We also thank M.\ Smith, V.\ Pota, C.\ Usher,  S.\ Kartha, N.\ Pastorello and J.\ Arnold for support during the preparation of this manuscript. The data presented herein were obtained at the W.M. Keck Observatory, which is operated as a scientific partnership among the California Institute of Technology, the University of California and the National Aeronautics and Space Administration. The Observatory was made possible by the generous financial support of the W.M. Keck Foundation. The analysis pipeline used to reduce the DEIMOS data was developed at UC Berkeley with support from NSF grant AST-0071048. J.P.B. and A.J.R. acknowledge support from the NSF through grants AST-0808099 and AST-0909237. CF acknowledges co-funding under the Marie Curie Actions of the European Commission (FP7-COFUND).

%
%
\bibliographystyle{mn2e}
\bibliography{ref}

\begin{appendix}
\section{Total subpopulation kinematic fits}

\begin{table*}\centering
\caption{Results for kinematic fits to various divisions of the three GC subpopulations. We tabulate the rotation velocity ($V_{rot}$), velocity dispersion ($\sigma$) and kinematic position angle (PA$_{kin}$) for each subpopulation in all the split definitions. Results are shown where the axis ratio ($q_{kin}=q_{phot}$) is fixed to the photometric values for each subpopulation ($q_{blue}=0.56$, $q_{green}=0.70$, $q_{red}=0.97$ and $q_{gal}=0.75$) and where it is fixed to 1. Numbers of GCs in each subpopulation representative sample are given in brackets after the definition parameters. The major axis of the NGC 4365 stellar light is $42^\circ$ or $222^\circ$. Probability is denoted with $P_{Colour}$}
\begin{tabular}{ccccccccccc}\hline
Split & Defn & $V_{rot}$ & $\sigma$ & PA$_{kin}$ & $V_{rot}$ & $\sigma$ & PA$_{kin}$ & $V_{rot}$ & $\sigma$ & PA$_{kin}$ \\
Definition & of $q_{kin}$ & (km s$^{-1}$) & (km s$^{-1}$) & ($^\circ$) & (km s$^{-1}$) & (km s$^{-1}$) & ($^\circ$) & (km s$^{-1}$) & (km s$^{-1}$) & ($^\circ$) \\ \hline
& & \multicolumn{3}{c}{$P_{Blue}>0.8$ (53)} & \multicolumn{3}{c}{$P_{Green}>0.8$ (53)} & \multicolumn{3}{c}{$P_{Red}>0.8$ (65)} \\ \hline
& $q_{phot}$ & $58\pm^{52}_{23}$ & $178\pm^{10}_{17}$ & $243\pm^{24}_{18}$ & $103\pm^{57}_{41}$ & $228\pm^{12}_{20}$ & $141\pm^{21}_{15}$ & $39\pm^{56}_{13}$ & $218\pm^{14}_{23}$ & $175\pm^{66}_{51}$ \\ 
med. prob. & & & & & & & & & & \\
& 1 & $49\pm^{48}_{18}$ & $179\pm^{10}_{17}$ & $254\pm^{38}_{54}$ & $83\pm^{57}_{29}$ &  $229\pm^{13}_{22}$ & $145\pm^{33}_{31}$ & $39\pm^{52}_{11}$ &  $218\pm^{14}_{22}$ & $174\pm^{62}_{53}$ \\ \hline
& & \multicolumn{3}{c}{$g'-i'<0.9$ (80)} & \multicolumn{3}{c}{$0.9<g'-i'<1.05$ (77)} & \multicolumn{3}{c}{$g'-i'>1.05$ (79)} \\ \hline
& $q_{phot}$ & $56\pm^{41}_{22}$ & $188\pm^{10}_{15}$ & $251\pm^{12}_{11}$ & $96\pm^{43}_{33}$ & $217\pm^{9}_{17}$ & $142\pm^{25}_{16}$ & $43\pm^{45}_{16}$ & $216\pm^{12}_{18}$ & $167\pm^{51}_{44}$ \\ 
col. cut 1 & & & & & & & & & & \\
& 1 & $40\pm^{41}_{18}$ & $189\pm^{9}_{15}$ & $267\pm^{40}_{50}$ & $\mathbf{81\pm^{39}_{24}}$ &  $\mathbf{217\pm^{10}_{15}}$ & $\mathbf{144\pm^{27}_{23}}$ & $43\pm^{45}_{15}$ &  $216\pm^{13}_{19}$ & $167\pm^{51}_{40}$ \\ \hline
& & \multicolumn{3}{c}{$g'-i'<0.9$ (80)} & \multicolumn{3}{c}{$0.9<g'-i'<1.1$ (100)} & \multicolumn{3}{c}{$g'-i'>1.1$ (56)} \\ \hline
& $q_{phot}$ & $56\pm^{43}_{21}$ & $188\pm^{9}_{14}$ & $251\pm^{17}_{12}$ & $81\pm^{43}_{29}$ & $221\pm^{9}_{13}$ & $126\pm^{17}_{14}$ & $95\pm^{50}_{32}$ & $202\pm^{15}_{25}$ & $195\pm^{23}_{21}$ \\ 
col. cut 2 & & & & & & & & & & \\
& 1 & $40\pm^{46}_{20}$ & $189\pm^{8}_{15}$ & $267\pm^{43}_{55}$ & $67\pm^{40}_{26}$ &  $222\pm^{9}_{14}$ & $124\pm^{28}_{24}$ & $\mathbf{94\pm^{48}_{34}}$ &  $\mathbf{202\pm^{15}_{26}}$ & $\mathbf{194\pm^{23}_{24}}$ \\ \hline
& & \multicolumn{3}{c}{$g'-i'<0.85$ (61)} & \multicolumn{3}{c}{$0.85<g'-i'<1.05$ (96)} & \multicolumn{3}{c}{$g'-i'>1.05$ (79)} \\ \hline
& $q_{phot}$ & $70\pm^{47}_{25}$ & $178\pm^{9}_{16}$ & $248\pm^{27}_{13}$ & $83\pm^{41}_{30}$ & $216\pm^{10}_{14}$ & $134\pm^{16}_{14}$ & $43\pm^{44}_{19}$ & $216\pm^{12}_{19}$ & $167\pm^{49}_{48}$ \\ 
col. cut 3 & & & & & & & & & & \\
& 1 & $\mathbf{67\pm^{44}_{27}}$ & $\mathbf{178\pm^{10}_{16}}$ & $\mathbf{269\pm^{28}_{34}}$ & $67\pm^{36}_{23}$ &  $217\pm^{10}_{15}$ & $134\pm^{31}_{25}$ & $43\pm^{45}_{15}$ &  $216\pm^{12}_{20}$ & $167\pm^{57}_{44}$ \\ \hline
\end{tabular}
\label{tab:kinall}
\end{table*}

The colour peaks and widths of the blue, green and red subpopulations cause significant overlap in the colour ranges of the subpopulations. We explore the effect of splitting the populations in different ways to determine the best way to minimize the contamination between samples of the subpopulations and maximize the number of GCs in each subpopulation sample.

The first subpopulation split definition is based on the probability that each GC is assigned of being in either the blue, green or red subpopulations by \citet{Blom12}. They assigned each GC a probability of belonging to the blue, green and red subpopulations ($P_{Blue}+P_{Green}+P_{red}=1$) based on its colour and galactocentric radius. We define medium probability samples with $P>0.80$ to represent the three subpopulations and find 53 blue, 53 green and 65 red GCs with line-of-sight radial velocity measurements. These samples are mostly free of contamination from the other subpopulations but have a relatively small number of GCs to be analysed for each subpopulation.

The next subpopulation split definitions cut the GCs into three subsamples at different $g'-i'$ colours. For the first colour cut split definition we split the sample at $g'-i'=0.9$ and $g'-i'=1.1$ where we see dips in number of the photometric sample colour distribution (see Fig. \ref{fig:hist}). This subpopulation split definition divides the kinematic sample roughly equally between the three subpopulations. The next two colour cut split definitions widen the green sample definition first to redder colours (colour cut 2) and secondly to bluer colours (colour cut 3). The repeated fits for the blue (in colour cut 1 and colour cut 2) and red samples (in colour cut 1 and colour cut 3) show the level to which results are affected by the randomisation in the bootstrap technique of determining uncertainties.

We also show the kinematic fits for both cases of the kinematic axis ratio. In the first case $q_{kin}$ is fixed to the photometric value for each GC subpopulation \citep[$q_{kin}=q_{photom}$ from][]{Blom12} and in the second the axis ratio is fixed to unity ($q_{kin}=1$). We do not find any significant differences in the fitted kinematics between the two axis ratio cases and conclude that this kinematic fitting method is not very sensitive to the kinematic axis ratio. In all split definitions for all subpopulations, the kinematic fits for the two cases are consistent within $1\sigma$ uncertainty and the computed $\Lambda/ndf$ values are equal. All further kinematic fits are done in the case where $q_{kin}=1$ as this is the more general case.

The best kinematic fits are found for the blue GCs in colour cut 3, green GCs in colour cut 1 and red GCs in colour cut 2. We determine this best kinematic fit to be the subpopulation sample has the most constrained position angle (for q=1), see Table \ref{tab:kinall}, and lowest value of $\Lambda/ndf$, see Table \ref{tab:min}. The blue subpopulation representative sample in colour cut 3 $\Lambda/ndf$ is marginally larger than in colour cut 1 or 2 but the position angle range is significantly smaller ($62^\circ$) than in either colour cut 1 or 2 ($90^\circ$ and $98^\circ$ respectively). The green subpopulation representative sample position angle range in colour cut 1 ($50^\circ$) is only slightly smaller than that of colour cut 2 ($52^\circ$) but the $\Lambda/ndf$ is significantly smaller. The red subpopulation representative sample $\Lambda/ndf$ is the same for each colour cut but the position angle range is significantly smaller in colour cut 2 ($47^\circ$) than either of the other colour cuts ($91^\circ$ and $101^\circ$) where the PA$_{kin}$ is not constrained to $<90^\circ$.

\begin{table}
\caption{Minimisation parameter divided by degrees of freedom ($\Lambda/ndf$) for kinematic fits to various divisions of the three GC subpopulations. Values for the best colour cuts are highlighted.}
\begin{tabular}{lccc}\hline
Split & \multirow{2}{*}{Blue} & \multirow{2}{*}{Green} & \multirow{2}{*}{Red} \\
Definition & & & \\ \hline
medium probability & 12.1 & 12.6 & 12.3 \\
colour cut 1 & 11.9 & \textbf{11.8} & 12.2 \\
colour cut 2 & 11.9 & 15.3 & \textbf{12.2} \\
colour cut 3 & \textbf{12.0} & 19.5 & 12.2 \\ \hline
\end{tabular}
\label{tab:min}
\end{table}

\section{Radial kinematic fits with fixed position angle}

\begin{figure}\centering
 \includegraphics[width=0.49\textwidth]{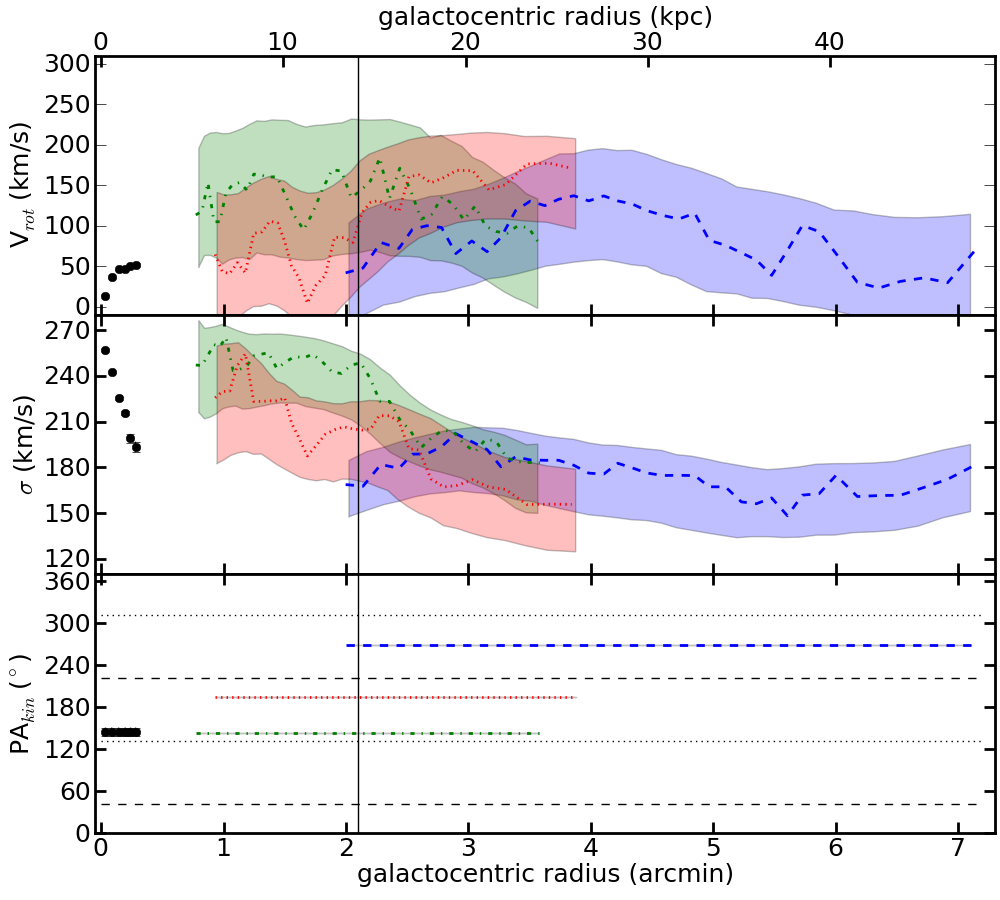}
 \caption{Kinematics as a function of radius for NGC 4365's three GC subpopulations. GC subpopulations are defined using the best colour cuts highlighted in Table \ref{tab:kinall} and the PA$_{kin}$ is fixed to the values found there. Symbols are as in Fig. \ref{fig:radkin}.}
 \label{fig:radkinPA}
\end{figure}

Here we compare the radial kinematic fits for the three subpopulations in the case where the PA is fixed to the value fitted for the whole subpopulation (Fig. \ref{fig:radkinPA}) to the case where the PA is allowed to vary with radius (Fig. \ref{fig:radkin}). The fitted values for $\sigma$ do not change at all when the position angle of rotation is fixed. The uncertainties in $V_{rot}$, are larger when PA$_{kin}$ is fixed than when the PA$_{kin}$ is allowed to vary. If the reduction in number of free parameters, in the case where PA$_{kin}$ is fixed, does allow a more stable fit to the $V_{rot}$ and $\sigma$ we would expect the uncertainties (68 per cent confidence intervals) to be smaller there. Since the opposite is true it is more likely that the fit is better where the PA$_{kin}$ is allowed to vary, and since the PA$_{kin}$ is well constrained at almost all radii in that case it is also likely that the $V_{rot}$ is also more reliable than the case where PA$_{kin}$ is fixed. 

The one exception is for red GCs less than 2 arcmin from the galactic centre. The varying PA$_{kin}$ is not well constrained there and rotation parameter ($V_{rot}/\sigma$) fails the criterion in Equation \ref{eq:vbias} \citep{St11}. In the case where PA$_{kin}$ is fixed $V_{rot}$ is correctly fitted to be consistent with zero. The decrease in $V_{rot}$ at certain other points in Fig. \ref{fig:radkinPA} (e.g.\ $\sim 3$ arcmin for the blue subpopulation) is due to the offset between the value the PA$_{kin}$ was set at and the PA$_{kin}$ at which the system is actually rotating at that radius.

\end{appendix}
\end{document}